\documentclass[usenatbib,letterpaper]{mn2e}
\usepackage{times}
\usepackage{graphicx}
\newcommand{\kpc}{\,\mbox{kpc}}
\newcommand{\kms}{\,\mbox{km}\,\mbox{s}^{-1}}
\newcommand{\msun}{\,{\rm M}_{\sun}}
\newcommand{\lsun}{\,L_{\sun}}

\newcommand{\tsim}{\sim\!}
\newcommand{\ea}{et al.~}
\newcommand{\gtrsim}{\ga} 
\newcommand{\lesssim}{\la} 

\title[Investigating the Andromeda Stream -- II]
{Investigating the Andromeda Stream: II. Orbital 
Fits and Properties of the Progenitor}
\author[M. A. Fardal et al.]
{M. A. Fardal$^{1,2}$\thanks{E-mail: fardal@fcrao1.astro.umass.edu,
babul@uvic.ca, jgeehan@uvic.ca, raja@ucolick.org}, 
A. Babul$^1$, J. J. Geehan$^1$, and P. Guhathakurta$^3$\\
  $^{1}$Dept. of Physics \& Astronomy, University of Victoria, 
  Elliott Building, 3800 Finnerty Rd., Victoria, BC, V8P 1A1, Canada\\
  $^2$Dept.\ of Astronomy, University of Massachusetts, 
      Amherst, MA, 01003, USA\\
  $^3$UCO/Lick Observatory, Dept.\ of Astronomy \& Astrophysics,
      Univ. of California, 1156 High St., Santa Cruz, CA, 95064, USA}
\date{Accepted 2005 November 10. Received 2005 October 30; in original form 2005 January 13}
\pagerange{\pageref{firstpage}--\pageref{lastpage}} \pubyear{2005}
\begin{document}
\maketitle
\label{firstpage}
\begin{abstract}
  We construct test-particle orbits and simple N-body models that
  match the properties of the giant stellar stream observed to the
  south of M31, using the model of M31's potential derived in the
  companion paper by Geehan \ea (2005).  We introduce a simple
  approximation to account for the difference in position between the
  stream and the orbit of the progenitor; this significantly affects
  the best-fitting orbits.  The progenitor orbits we derive have orbital
  apocenter $\tsim 60 \kpc$ and pericenter $\tsim 3 \kpc$, though
  these quantities vary somewhat with the current orbital phase of the
  progenitor which is as yet unknown.  Our best combined fit to the
  stream and galaxy properties implies a mass within 125 kpc of M31 of
  $(7.4 \pm 1.2) \times 10^{11} \msun$.  Based on its length, width,
  luminosity, and velocity dispersion, we conclude that the stream
  originates from a progenitor satellite with mass $M_s \sim 10^9
  \msun$, and at most modest amounts of dark matter; the estimate of
  $M_s$ is again correlated with the phase of the progenitor.  M31
  displays a large number of faint features in its inner halo which
  may be progenitors or continuations of the stream.  While the
  orbital fits are not constrained enough for us to conclusively
  identify the progenitor, we can identify several plausible
  candidates, of which a feature in the planetary nebula distribution
  found by Merrett et al.\ is the most plausible, and 
  rule out several others.  We make predictions for the kinematic properties 
  of the successful candidates.  These may aid in observational
  identification of the progenitor object, which would greatly
  constrain the allowed models of the stream.
\end{abstract}
\begin{keywords}
galaxies: individual: M31 -- galaxies: interactions -- galaxies: kinematics and dynamics
\end{keywords}
\section{INTRODUCTION}
The giant southern stream observed near the Andromeda galaxy (M31)
provides us with an extraordinary opportunity to study the disruption
by a large disk galaxy of one of its satellite galaxies, and thereby 
obtain a sensitive probe of the dynamics in M31.
The stream was originally
discovered through counts of red giant stars in the M31 halo
\citep{ibata01}. Later extensions of this survey
\citep{ferguson02,mcconnachie03,ferguson04} outlined the extent of the
stream from a projected distance of 60 kpc from the center of M31
on the southern side, to
a possible extension on the northern side of M31.  
Fits to the tip of the red giant branch \citep{mcconnachie03} provided
line-of-sight distances as a function of position along the stream.
Finally, spectroscopy of stars along the path of the stream
\citep{ibata04,raja05} has resolved the stream as a distinct kinematic
component as well, and has also provided absorption-line estimates of
the metallicity of the stream stars.  The stream resembles the
extended tidal tails of the Sagittarius dwarf galaxy that wrap around
the Milky Way \citep{totten98}, though its azimuthal extent around M31 is 
much smaller.  A major difference is that we do not
know if the progenitor of the Andromeda stream has survived and if so
where it is, although several faint features in the inner halo of M31
have been suggested as possible progenitors.

Since the three-dimensional path of the stream is now measured
along with one velocity component, we have nearly complete information
on the path of the system in phase space.
If we combine the stream data with other observations of
M31 such as its rotation curve, we should be able to obtain a direct
estimate of the gravitational potential in the M31 halo, as well as
follow the path of the stream beyond the visible southern portion.
Current work includes some first steps in this direction.
\citet{font04} computed orbits in a bulge-disk-halo model.  
These authors found rough consistency of their
potential with the stream, but they did not optimize the agreement
in detail, focusing instead on the physical implications of the
width and velocity dispersion of the stream.  \citet{ibata04} tested
several potentials, deriving a weakly model-dependent limit on the
mass within 125 kpc.  They supplied model parameters only for
their simple spherical potentials, which have the disadvantage that
they do not match the M31 rotation curve.  They used a more
sophisticated bulge-disk-halo model as well, but this is not fully
specified in the paper.

An issue that deserves closer attention is the slight difference
between the locations of the orbit and the stream.  Both
\citet{ibata04} and \citet{font04} assume for simplicity that the
stream follows the orbit.  However, the stream exists precisely
because its constituent stars are on slightly {\em different} orbits.
The systematic variation of orbital energy along the stream, which is
accounted for implicitly by N-body simulations but not by simple
orbital integrations, may affect the estimates of the M31 potential and
the trajectory of the progenitor.  Hence it is worth building upon the
earlier work to construct a more specific and more accurate model of
the stream.

In \citet{geehan05} (Paper I), we presented a simple analytic
bulge-disk-halo model for the potential of M31, fitted to observed
tracers of the potential ranging from bulge stars to satellites in the
outer halo, and we quantified the error on the model parameters.  In
this paper, we use this fit to derive simple models of the southern
stream, and discuss their physical implications.  We start off with a
subset of the observational data and then gradually add in more
constraints.  In \S~2, we consider the orbital information alone.  We
integrate orbits in our M31 potential to obtain best values and
confidence intervals for the orbital parameters, and further refine
the galactic parameters compared to Paper I.  We introduce an
approximate method to calculate the phase space location of the 
stream given the orbit of the progenitor, test it with 
N-body simulations, and incorporate it within the
fitting procedure; as a result the allowed orbits are changed
considerably.  In \S~3, we add in the constraints that the observed
length, width, velocity dispersion, and luminosity of the stream put 
on the properties and current location of the progenitor satellite, 
using N-body simulations that follow our best-fitting orbit.  \S~4 examines 
the forward continuation of the stream, including the progenitor location,
and its relation to the orbital characteristics of the stream.  Among
the numerous faint features in the inner halo of M31, we can identify
some that are plausibly related to the stream.  Finally, \S~5 presents
our conclusions.
%
%
\section{Orbital fits}
\subsection{Summary of the observations}
Though it was discovered only recently, the variety of observational
information about the southern stream already approaches that which is
available for the Sagittarius stream in our own galaxy.  We review
here what is currently known about the configuration of M31 and the
southern stream.

We look at the M31 galaxy from below, as is evident from images of the
disk's dust lanes projected on the bulge, and the galaxy spins
counterclockwise on the sky.  We take the inclination of M31 to be
$77^\circ$, and the position angle to be $37^\circ$.  We use a
distance to M31 of $z_g = 784 \pm 24 \kpc$ \citep{stanek98}, and a
mean radial velocity of $v_{gz} = -300 \pm 4 \kms$ \citep{devauc91}.

Because the M31-stream system covers such a large region on the sky,
its transverse velocity is one of the primary uncertainties in its
internal kinematics, which is a situation rarely encountered in
astronomy.  Since M31 and the Milky Way are the primary mass concentrations
in the Local Group, it is traditionally assumed that they are falling
in toward each other for the first time, so that the transverse
velocity is much smaller than the radial velocity.  
We prefer to use the results of \citet{peebles01}, who
have modeled the Local Group dynamics in detail. They found solutions
for M31's transverse velocity falling mostly into two clumps in
opposing directions, with velocities in supergalactic coordinates $l$
and $b$ of either
$(v_{gl},v_{gb}) = (-40,120) \kms$ 
or $(30,-140) \kms$.
The eastward and northward velocities are then
$(v_{gx},v_{gy}) = (-130, 20)\kms$ 
or $(150,0) \kms$. The dominance of the east-west component stems from
the direction of the local filament, along which M31 tends to move in
the Peebles \ea model.
 
The giant southern stream extends to the SE of the galaxy for
approximately $5^\circ$.  The total extent of the stream is not well
known; at its southern end it {\it may} be visibly petering out close
to Field 1 of \citet{mcconnachie03}, as seen in the larger survey
presented in \citet{ferguson04}, and continuing northwards it is
eventually lost to confusion against the M31 disk.  The stream
luminosity is $\tsim 3.4 \times 10^7 \lsun$ \citep{ibata01},
suggesting a stellar mass $\tsim 2.4 \times 10^8 \msun$ if we assume
$M/L_V \approx 7$.  The mean metallicity of the stream is 
$\mbox{[Fe/H]} \approx -0.5$ \citep{raja05}, 
suggesting the progenitor had a much larger
total luminosity of $\tsim 10^9 \lsun$ \citep{font04}.  McConnachie
\ea suggest the stream continues at least $1.5^\circ$ to the NW on the
near side of M31, but this is questionable for several reasons: the
purported extension is not a distinct surface brightness feature
\citep{ferguson04},
there is as yet no spectroscopic confirmation of a stream component,
and as shown in \citet{ibata04} it is difficult to derive an orbit
that fits this feature.  We therefore consider this ``northern
extension'' to be just one of many faint features that could possibly
be extensions of the stream.  In contrast, the reality of the
southern portion is in no doubt, due to the strong enhancement in star
counts and observation of a distinct kinematic component in five
spectroscopic fields along the stream.  

Field coordinates for the published observations of the stream are
listed in Table~\ref{table.obs}, along with distances and velocities
of the stream.  The positions show
the stream is a roughly linear structure, at least in projection.  The
field ``a3'' of \citet{raja05} overlaps the stream only by chance, and is
not centered on the stream.  \citet{mcconnachie03} chose their fields
to lie approximately along the stream center.  However,
some of these positions were chosen by extrapolation from the inner
part of the stream.  In fact, their fig\. 3, showing a histogram of
stars as a function of transverse position, suggests the central ridge
of the stream is offset fairly uniformly by $\tsim 0.15^\circ$ from
the field centers, towards the SE.  We use the slope of Fields 1--8 to
define rotated sky coordinates $m$ and $n$, where $m \equiv 0.504 \xi
-0.864 \eta$ increases along the stream (roughly toward the SE) and $n
\equiv -0.864 \xi -0.504 \eta$ increases across the stream (roughly
toward the SW).  We then shift the mean transverse position of the
stream relative to the field centers by $0.15^\circ$.  This shift is
small compared to the width of the stream, approximately $0.5^\circ$
wide full width at half-maximum (FWHM)
on the sky or about 7 kpc in distance, as seen in the same
figure. We assign an uncertainty of $0.15^\circ$ to the stream's
angular position.

\begin{table}
\caption{Kinematic data for the southern stream, in units where M31 is
at the center.  A dash indicates that there is no data for that field.
The angular positions $\xi$ and $\eta$ are those
of the field centers; field ``a3'' is certainly offset from the position
of the stream, while fields 1--8 may be as well.}
\label{table.obs}
\begin{tabular}{@{}ccccc}
\hline
Field & $\xi$~(deg) & $\eta$~(deg) & $d'$~(kpc) & 
$v'_{r}$~(km~s$^{-1}$)  \\
\hline
a3 & $+1.077$ & $-2.021$ & 66 & $-158$ \\
\hline
1 & $+2.015$ & $-3.965$ &  102  & 0 \\
2 & $+1.745$ & $-3.525$ &   93  & $-50$ \\
3 & $+1.483$ & $-3.087$ &   76  & --- \\
4 & $+1.226$ & $-2.653$ &   71  & --- \\
5 & $+0.969$ & $-2.264$ &   56  & --- \\
6 & $+0.717$ & $-1.768$ &   52  & $-180$ \\
7 & $+0.467$ & $-1.327$ &   45  & --- \\
8 & $+0.219$ & $-0.886$ &   $-4$  & $-300$ \\
12 & $-0.731$ & $+0.891$ & $-45$  & --- \\
13 & $-0.963$ & $+1.342$ & $-26$  & --- \\
\hline
\end{tabular}
\end{table}

\begin{table*}
\begin{minipage}{175mm}
\caption{Parameter choices for some orbits discussed in the text.
The first six parameters give the initial position and velocity
of the test particle (the orbit is calculated both forward and
backward of this point).  $\Delta \! z$ is the fitted systematic offset
between our distance calibration and that of \protect\citet{mcconnachie03}. 
The halo parameter $f_h$ determines the strength of the galaxy potential.  
The final parameter gives the value of reduced $\chi^2$ from our fit.
Initial conditions 1e, 2e, 3e, and 4e give the same orbits as 
1, 2, 3, and 4, except they begin at an earlier point on the orbit.}
\label{table.streamfits}
\begin{tabular}{rrrrrrrrrrrl}
\hline\\  
 Orbit  & $x(0)$  &  $y(0)$  &  $z(0)$  
            &  $v_x(0)$  &  $v_y(0)$  &  $v_z(0)$  
 &  $\Delta \! z$  &  $f_h$  &  $F_p$  &  $\chi^2 / N_{deg}$  & comments  \\
            &  $\kpc$  &  $\kpc$  &  $\kpc$  
            &  $\kms$  &  $\kms$  &  $\kms$  
 &  $\kpc$      &         &       &   \\
\hline
\hline\\
  1~~~~& 5.19    &  $-20$   &  38.0
    &  $-61.2$  &  132.9  &  $-214.4$
    &  $-4.0$   &  $1.14$ & ... &   0.16  &  fit to orbit, variable halo \\
    ~~~~& $\pm 1.12$    &  ...   &  $\pm  8.8$
    &  $\pm   9.4$  &  $\pm  21.9$  &  $\pm    6.7$
    &  $\pm 12.8$   & ... &  $\pm 0.35$ &         &  \\
  1e~~~& $-52.7$  &  $42.6$   &  -28.2
    &  $126.5$  &  $-94.2$  &  $53.9$
    &  $-4.0$   &  $1.14$ & ... &   0.16  &  same orbit, earlier start \\
  2~~~~&  5.60    &  $-20$   &  18.4
    &  $-67.9$  &  145.2  &  $-209.5$
    &  $-21.42$ & $ 1.14$ & $1.00$ & 0.51 
    & fit to stream, var halo, $F_p = 1.0$  \\
    ~~~~& $\pm  1.02$    &  ...   &  $\pm  2.7$
    &  $\pm   4.0$  &  $\pm  13.7$  &  $\pm    7.3$
    &  $\pm   6.4$   &  $\pm  0.37$   & ... &         &  \\
  2e~~~&   $-8.8$       &  $-19.1$    &   $-56.4$   
    &  4.9  &  44.8  &  42.4
    &  $-21.42$ & $ 1.14$ & $1.00$ & 0.51 & same orbit, earlier \\
  3~~~~&  7.04    &  $-20$   &  23.6
    &  $-51.9$  &   85.2  &  $-155.1$
    &  $-18.70$   &  $ 1.16$ &  $1.50$ &    0.48  &  
    fit to stream, var halo, $F_p = 1.5$  \\
    ~~~~& $\pm  0.66$    &  ...   &  $\pm  5.8$
    &  $\pm   6.5$  &  $\pm  14.3$  &  $\pm    8.5$
    &  $\pm  11.4$   &  $\pm  0.40$   &  &         &  \\
  3e~~~&   $-5.5$       &  $-11.8$        &   $-42.1$   
    &  5.9 &  45.7 &  53.2
    &  $-18.70$   &  $ 1.16$ &  $1.50$ &    0.48  & same orbit, earlier \\
  4~~~~&  5.15    &  $-15$   &  16.7
   &  $-59.4$  &  102.3  &  $-175.0$
   &  $-19.88$   &  $ 1.10$ & $2.00$ & 0.46  
   &  fit to stream, var halo, $F_p = 2.0$  \\
   ~~~~& $\pm  0.51$    &  ...   &  $\pm  2.8$
   &  $\pm   4.1$  &  $\pm  12.0$  &  $\pm    7.8$
   &  $\pm   8.0$   &  $\pm  0.38$   &  $\pm  0.00$ &         &  \\ 
  4e~~~&   $-2.2$       &  $-7.5$        &   $-32.2$   
   &  4.4 &  52.8 &  81.1
   &  $-19.88$   &  $ 1.10$ & $2.00$ & 0.46  & same orbit, earlier \\
  5~~~~&  8.54    &  $-20$   &  29.4
   &  $ -7.2$  &    6.4  &  $ -57.7$
   &  $-15.55$   &  $ 1.02$ &    $2.02$ &    0.97  &  fit to M32 \\
   ~~~~& $\pm  0.60$    &  ...   &  $\pm  2.4$
   &  $\pm  17.9$  &  $\pm  28.1$  &  $\pm   54.5$
   &  $\pm   5.1$   &  $\pm  0.32$   &  $\pm  0.01$ &         &  \\
  6~~~~&  5.78    &  $-20$   &  25.7
   &  $-62.4$  &  118.9  &  $-196.7$
   &  $-12.86$   &  $ 1.47$ &    1.12 &    0.57  
   & fit to Northern Spur \\
   ~~~~& $\pm  0.93$    &  ...   &  $\pm  3.3$
   &  $\pm   7.4$  &  $\pm  16.8$  &  $\pm    9.3$
   &  $\pm   8.1$   &  $\pm  0.46$   &  $\pm  0.03$ &         &  \\
  7~~~~&  6.96    &  $-20$   &  25.1
   &  $-51.0$  &   83.2  &  $-154.5$
   &  $-16.20$   &  $ 1.18$ &    1.50 &    0.47  &  fit to And NE \\
   ~~~~& $\pm  0.71$    &  ...   &  $\pm  2.8$
   &  $\pm   9.4$  &  $\pm  17.9$  &  $\pm   12.7$
   &  $\pm   6.4$   &  $\pm  0.40$   &  $\pm  0.13$ &         &  \\
  8~~~~&  5.25    &  $-15$   &  24.2   
   &  $-63.6$  &  136.4  &  $-220.9$
   &  $ -7.07$   &  $ 2.21$ &    1.29 &    1.00  &  
   fit to Merrett NE PNe\\
   ~~~~& $\pm  0.52$    &  ...   &  $\pm  3.2$ 
   &  $\pm   8.0$  &  $\pm  19.2$  &  $\pm    7.7$
   &  $\pm   9.1$   &  $\pm  0.68$   &  $\pm  0.04$ &         &  \\
\hline
\end{tabular}
\end{minipage}
\end{table*}
Distances to the M31 stream fields are given by \citet{mcconnachie03}
and reproduced in Table~\ref{table.obs}.  These show the stream lies
further away than M31, and the stream follows a roughly radial trajectory.  
We exclude the fields on the northern side, and exclude the field nearest
to M31 (Field 8) as well because of possible contamination by disk
stars.  McConnachie \ea estimate a total error of 20 kpc on their
distances, but this is composed of $\tsim 12 \kpc$ random error and
$\tsim 16 \kpc$ systematic error from the photometric zero-point
and the stellar population uncertainties.  Orbits that have a steeper
or shallower gradient of distance along the stream than the observed
gradient, as in fig.\ 4c of \citet{ibata04} or fig.\ 2a of \citet{font04}, 
can thus be ruled out even if they lie within the total error bars of
McConnachie et al.  To treat this in our orbital fitting, we take the
observed distances to be the McConnachie values plus a free parameter,
$\Delta \! z$.  Assuming the uncertainty on this systematic shift is
$\sigma_{\Delta \! z} = 16 \kpc$, we then include a term $(\Delta \!
z/\sigma_{\Delta \! z})^2$ in the total $\chi^2$.  We set the {\em
  random} error on each distance point equal to $12 \kpc$.

The mean velocity at various points along the stream has been measured
in four fields along the stream by \citet{ibata04}, and in the single
stream field ``a3'' of \citet{raja05}.  This reveals that the stream is
falling in toward M31 from behind.  We again omit the field closest to
M31 (Field 8 of Ibata et al.)  because of contamination concerns.
\citet{raja05} estimate the random error on their mean velocity to be
$4 \kms$. \citet{ibata04} do not quote errors on their mean
velocities, but they are probably comparable since they estimate the
velocity dispersion itself to be only $11 \kms$.  However, the true
uncertainty includes systematic effects such as the $4 \kms$ error in
M31's radial velocity, and contamination by the possibly lumpy velocity
distribution of the general halo population.  Because of these
concerns, we (somewhat arbitrarily) 
raise the estimate of the velocity error to $8 \kms$.

Our orbits are computed in a rectangular right-handed
coordinate system $(x,y,z)$,
centered on M31 and oriented with respect to the sky so that 
$x$ points east, $y$ points north, and $z$
points away from us.  However, the observed quantities are not
obtained in a rectangular coordinate system, because M31 is so close
that the system has a significant width and depth on the sky.
Insufficient information is available to correct all the observations
to rectangular coordinates, so we convert the orbits to observed
coordinates instead.  The positional coordinates $\xi$ and $\eta$ are
standard coordinates on the tangent plane.  We obtain the east-west
coordinate $\xi$, the north-south coordinate $\eta$, the distance $d$,
and the radial velocity $v_r$ using the following formulae:
\begin{eqnarray}
\xi \! & \! = \! & \! x/(z_g + z)\\
\eta \! & \! = \! & \! y/(z_g + z)\\
d \! & \! \equiv \! & \! d' + z_g = \left[ x^2 + y^2 + (z_g + z)^2 \right]^{1/2}\\
v_r \! & \! = & d^{-1} \left[ x (v_x\! +\! v_{gx}) + y (v_y \! + \! v_{gy}) 
   + (z_g \! + \! z) (v_z\! +\! v_{gz}) \right]  \\
  & \equiv & v'_r + v_{gz} \nonumber
\end{eqnarray}
In our plots we use the relative distance $d'$ (which we call the
``depth'') and the relative velocity $v_r'$, found by subtracting the
mean distance and radial velocity of M31.  These equations could also
be expanded in orders of M31-centric quantities, since in practice
only the first-order correction terms are significant.  For the two
solutions for the transverse velocity of M31 $(v_{gx}, v_{gy})$
discussed above, the effect of M31's transverse velocity
on $v_r$ is quite modest, only about $\pm 5 \kms$ for Field 1 which is
the most strongly affected field.  This is partly because these
velocity solutions point almost perpendicular to the stream.  In
contrast, a maximum velocity of $\pm 300 \kms$ along the stream, as
allowed for by \citet{ibata04}, would have added a significant
uncertainty of $\tsim 30 \kms$.  We set $v_{gx} = v_{gy} = 0$ from now
on.
\subsection{Fits to stream directly}
\begin{figure*}
\includegraphics[width=16cm,bb=0 8 504 495]{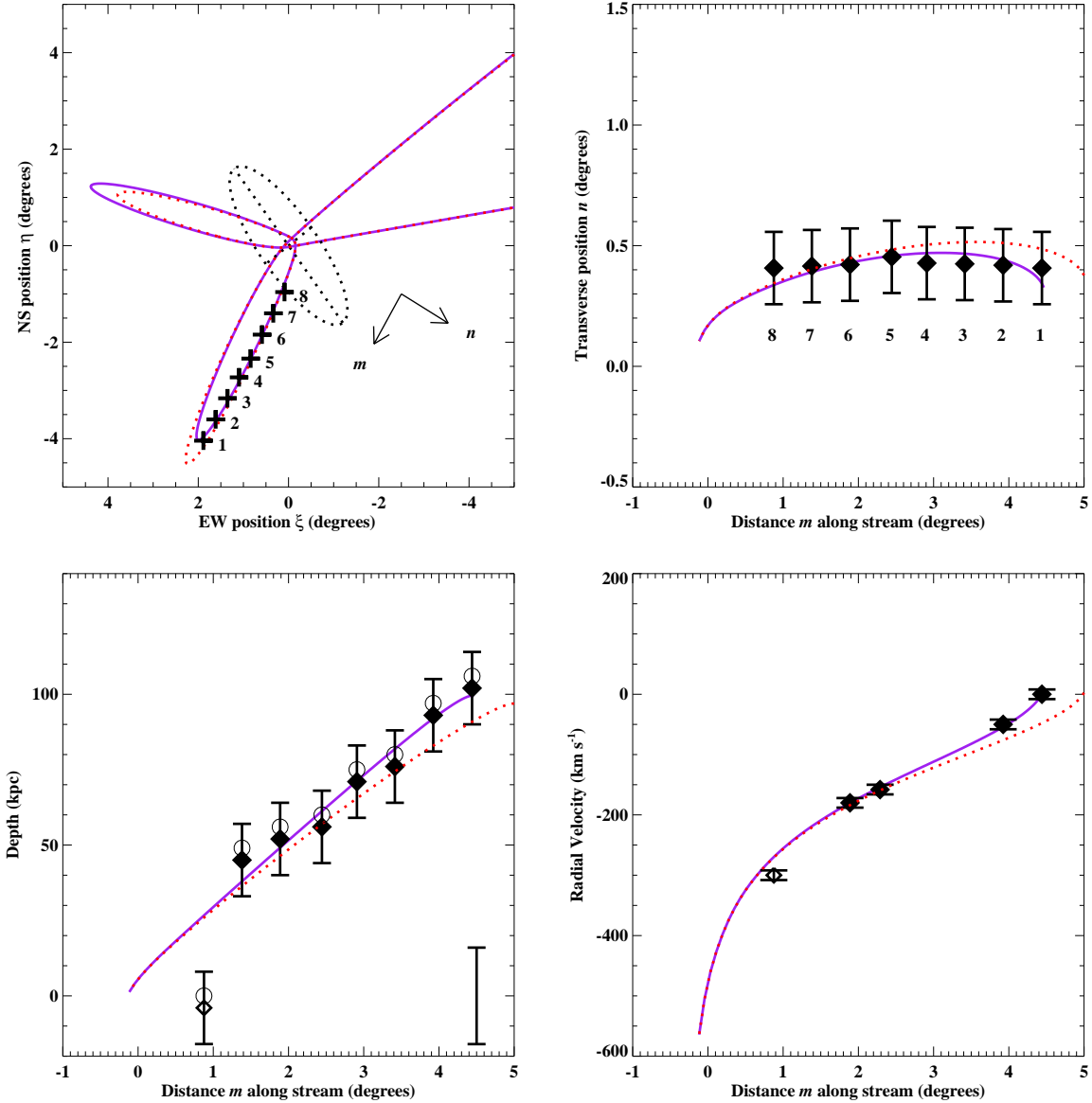}
\caption{
\label{fig.orbit}
Best-fitting orbit assuming the stream follows the orbit, using the
flattened disk potential. The orbit is shown with purple solid lines.
The orbit calculated using a rectilinear coordinate system is shown
with red dotted lines (see discussion in the text).  Observational
results are shown with black symbols.  (Upper left): North-south
coordinate $\eta$ versus east-west coordinate $\xi$.  The direction
vectors of the stream coordinates $m$ and $n$ are shown in the plot.
Crosses indicate the stream positions derived from the field
coordinates of \protect\citet{mcconnachie03}, offset slightly as
discussed in the text; the fields are labeled according to the numbering
in that paper.  (Upper right): Transverse stream coordinate
$n$, versus distance along stream $m$, with the vertical scale
expanded for clarity.  Diamonds indicate the same stream positions as
in the previous panel, again labeled by field number.  
(lower left): Depth $d'$ relative to M31
versus $m$.  Diamonds indicate the observational constraints of
\protect\citet{mcconnachie03}, as adjusted by the systematic shift
$\Delta \!  z$ from our fit; open circles give the depth values with
the original calibration.  The leftmost diamond is left empty as a
reminder we do not use this point in the fit, due to concerns over
contamination by the disk.  The error bar in the bottom right corner
indicates the uncertainty on $\Delta \! z$.  (Lower right): Radial
velocity $v_r'$ relative to M31, versus $m$.  Diamonds indicate the
observational constraints of \protect\citet{ibata04} and
\protect\citet{raja05}.  The leftmost symbol is again left empty as a
reminder we do not use this point in the fit.}
\end{figure*}
We now try to find orbits that fit the stream observations.  At first,
for simplicity, we assume that the stream follows the orbit, as in the
work of \citet{ibata04} and \citet{font04}.  In Paper I, we
demonstrated that even with this assumption our orbits can still
differ significantly from those in the cited papers, because of the
different radial mass distribution and the flattening of the disk
potential.  The possible flattening of the halo component is a
contentious issue even in the case of the Sagittarius stream in our
own Galaxy 
\citep{ibata01sag,martinez04,johnston05,law05},
which is more suitable for detecting the effects of a non-spherical
potential due to its much larger azimuthal extent.  In any case the
disk contributes significantly to the overall flattening in 
the Milky Way, and thus probably in M31 as well.  In this paper we
will simply assume the halo is spherical, but this is certainly an
issue to examine in the future.

We ignore the effect of dynamical friction on the satellite orbit.
This is partly for convenience, as the strength of dynamical friction
depends on the unknown mass and size of the progenitor, and thus
complicates the estimates we make of these quantities in \S~3 below.
More importantly, we expect the effect of dynamical friction on the
energy to be fairly small ($<10\%$ per orbit) unless the progenitor
satellite mass $M_s$ is quite large ($M_s \gg 10^9 \msun$).  Our
simulations typically have $M_s \sim 10^9 \msun$ at the start of the
run, and the effects of tidal stripping and gravitational disk
shocking further reduce this mass during the run.  Dynamical friction
due to the disk itself is likely important only for satellites with
orbits in the plane of the disk \citep{taylor01}.  Furthermore, for a
highly radial orbit such as that required by the southern stream, the
energy loss is almost entirely due to the passage through pericenter.
The energy within the stream, which is observed over less than one
radial period, is thus affected very little by the neglect of
dynamical friction.  In \S~4, we speculate on the location of the
progenitor allowing for the possibility that it has passed through
pericenter once again.  The neglect of dynamical friction there may
induce a small error, the size of which depends on how much of the
mass has managed to remain as an intact object during this second
passage.  Accounting for this would greatly complicate the analysis,
so we leave treatment of dynamical friction to future work.

We also assume a static potential, ignoring any interaction with M31's
satellite galaxies or recoil and/or distortion of M31's potential.
The latter assumption is justified because of the large mass of
M31 ($\tsim 10^{11} \msun$ in the bulge and disk alone), compared to
the typical estimated mass of the progenitor ($\tsim 10^9 \msun$) that
we estimate below.  With these assumptions, the zero-point of the
orbit is arbitrary.  We absorb this freedom by setting the orbit to
begin in a fixed plane, usually $y(0) = -20 \kpc$.

Paper I showed that there is considerable uncertainty in the potential
of M31, especially in the halo regions where the stream lies.  This
uncertainty is mostly due to two degeneracies between the model
parameters, namely a tradeoff between the disk and halo components and
between the scale radius $r_h$ and density parameter $\delta_c$ of the
halo.  Much of the formally allowed region, however, implies unlikely
values for the physical parameters like the halo concentration,
baryonic fraction, and disk mass-to-light ratio; 
thus the likely region of parameter space is much
smaller.  Because the detailed properties of the disk and bulge are
not probed by the stream, we limit our exploration of parameter space
to a one-dimensional path that varies the strength of the force at
large radii.  We set the disk surface density to a fixed value
$\Sigma_0 = 4 \times 10^8 \msun \kpc^{-2}$; the bulge and disk scale
lengths $r_b = 0.61 \kpc$ and $r_d = 5.4 \kpc$ are already fixed by
our fit to the surface brightness.  (These parameters are defined
completely in Paper I.)  We then find the best fit values
of the bulge mass $M_b$ and the halo density parameter $\delta_c$ as a
function of the halo scale radius $r_h$, storing them in a lookup
table along with the value of $\chi^2$.  In other words, we are
probing a vertical path through the contours in fig.\ 6 of Paper I.
For convenience, we describe the location along this path
with a ``halo parameter'' $f_h \equiv r_h / r_h^{(0)}$, 
where $r_h^{(0)} = 9.0 \kpc$ is the best fit scale radius
for the chosen $\Sigma_0$ in our flattened disk model.  To within a
few percent, the best-fitting bulge mass and halo density parameter 
along this path are given by
\begin{eqnarray*}
M_b & = & (2.71 + 0.435 u - 0.161 u^2 + 0.0229 u^3) \times 10^{10} \msun \; ,\\
\delta_c &= & \exp(12.46 - 1.928 u + 0.1434 u^2) \; ,
\end{eqnarray*}
with $u \equiv \ln f_h$.  The mass within 125 kpc, approximately
the maximum radius of the stream, scales roughly as $M_{125} = 6.68
\times 10^{11} f_h^{0.57} \msun$.  The allowed deviation of $\delta_c$ 
from its best-fitting value is
very small once the other parameters are specified, due to the small
error bars on the rotation curve, so we do not try to adjust it from
its best fit value to vary the halo potential further.

This gives seven free parameters to consider: $x(0)$, $z(0)$, $v_x(0)$,
$v_y(0)$, $v_z(0)$, $\Delta \! z$, and $f_h$.  $x(0)$ and the ratio
$v_x(0) / v_y(0)$ are fairly tightly constrained by the positional
data, while $v_z(0)$ is constrained by the spectroscopic data.  This
leaves $z(0)$ and $v_x(0) / v_z(0)$ as the most uncertain parameters,
since they involve the distance measurements.  

We use a simple Runge-Kutta integration procedure to trace the orbit,
since extremely high accuracy over many orbital periods is not
required.  We use a Levenberg-Marquardt $\chi^2$ routine to optimize
the fit to the stream.  

Our best-fitting orbit (labeled orbit 1 in Table~\ref{table.streamfits})
is shown as the solid line
in Fig.~\ref{fig.orbit}.  We also show the same orbit calculated
with a constant conversion of transverse distance to angle, using the
distance to M31 of 784 kpc.  When this is done, the stream moves outward
significantly in $\xi$ and $\eta$ because it lies further away than
M31. $v'_r$ also changes by a small amount ($\lesssim 5 \kms$) due to
the variation in the radial vector on the sky.

Our best-fitting orbit fits the data quite well; in fact, the $\chi^2 =
2.2$ for 14 degrees of freedom is almost too good.  (In part, this is
because the positional data points are derived from the field centers
rather than calibrated individually,
and thus have effectively been smoothed.)  The best-fitting
parameter values and their formal errors from the covariance matrix
are given in Table~\ref{table.streamfits}.  The radial period is 1.7
Gyr, apocenter is at 119 kpc, and pericenter is at 2.2 kpc.  
The orbit reaches apocenter near the most distant field (Field 1); this 
happens because the velocity relative to M31 is close to zero,
requiring turnaround there.  If we include only the observational
constraints from the stream, the preferred halo parameter is $f_h =
1.8 \pm 1.1$, corresponding to a mass of $M_{125} = (9.1 \pm 3.1)
\times 10^{11} \msun$.  This
suggests a more massive halo than in our preferred galaxy model, which
has $f_h=1.0 \pm 0.4$, or $M_{125} = (6.7 \pm 1.3) \times 10^{11}
\msun$.  The observational constraints on the galaxy fit are stronger
than those on the stream, however, so when we combine the stream data
with the galaxy fit the preferred halo scaling parameter is raised
only slightly: $f_h = 1.1 \pm 0.3$, or $M_{125} = (7.2 \pm 1.0) \times
10^{11} \msun$.  
The fits with and without the galaxy term are
both consistent with \citet{ibata04}, who found a $1\sigma$
confidence interval of 
$M_{125} = (6.2$--$10.0) \times 10^{11} \msun$.
The orbit is less strongly radial than in Ibata et al., 
which has $R_{apo} \sim 125 \kpc$ and $R_{peri} \sim 1.8 \kpc$.
It is more radial than the orbits in \citet{font04},
which have $R_{apo} \sim 110 \kpc$ and $R_{peri} \sim 4 \kpc$. 
The direction of their second loop is also significantly different.
The differences between these various orbits are induced 
both by different assumptions about the potential of M31 
and by differences in the chosen fitting technique and 
treatment of the observational data.
\subsection{Stream-orbit tilt}
The next step is to incorporate a phase space tilt between the orbit
and the stream.  When a satellite is tidally disrupted, stars on
different orbits arrange themselves primarily as a function of their
orbital energy onto a stream that nearly follows the orbit
\citep{johnston01}.  The ``nearly'' is important, though: the stream
is tilted relative to the progenitor's orbit, both in 3-d real space
and 6-d phase space.  For an object like the Andromeda stream which is
on a highly radial orbit, the stream stars are lost from the
progenitor almost entirely at pericenter.  If the progenitor is
disrupted over the course of several radial periods, a stream results
from each pericentric passage, with the youngest stream having the largest tilt
relative to the orbit.  (Fig.~1 in \citealp{law05} displays these
effects for a model of the Sagittarius stream.)  The length of the
stream increases both with time and with the scatter in the energy;
this scatter depends in turn on the satellite mass.  The phase space
tilt between the orbit and the stream, however, decreases with time
and to first order is {\em independent} of the satellite mass.  Hence,
to predict the stream location given the progenitor's orbit, we need
only to know when the stream stars were removed from the progenitor,
and not the mass or other properties of the progenitor.  In theory
this could include pericentric passages before the last one, but this
is unlikely because of the narrowness and high surface brightness of
the stream \citep{font04}.  (Note: when we refer to the ``last''
pericentric passage, we mean the last one experienced by the southern
stream stars, unless otherwise specified.  The progenitor itself may
already have passed through pericenter again.)

The current location of the progenitor along the orbit thus becomes an
important variable.  This is currently unknown, but we can at least
put some constraints on it.  The progenitor is unlikely to trail the
southern stream in its orbit, since there would be another stream
behind it which is not observed.  The progenitor is not observed along
the stream, although it could be located there if it is almost
entirely disrupted.  However, the luminosity of the progenitor
inferred from the stream's metallicity by \citet{font04} is much
larger than the total luminosity of the stream, suggesting this is rather
unlikely.  If the progenitor lies too far ahead of the stream in its
orbit, on the other hand, the corresponding stream length and hence
the implied scatter in the orbital energies may require a improbably
large progenitor mass, as discussed below.

Suggested progenitors for the stream include the And VIII
concentration, the Northern Spur, and a high-metallicity feature on
the east side of the disk.  All lie near the M31 disk, and thus would
have to precede the stream in its orbit.  If one of these was the
source of the stream, one might expect that if it had passed the next
pericenter and was heading inbound again, there would be an obvious
continuation of the stream pointing outwards.  The lack of such a
feature is a weak argument that the orbital phase of the progenitor is
$< \! 1.5$.  It is also possible
that the progenitor lies somewhat ahead of the southern stream in the
vicinity of pericenter, but is projected against M31 and has escaped
detection completely.  These arguments suggest the
stream stars were most likely unbound from the satellite in the range
0.8--1.5 radial periods ago, at the last pericenter passage.

To fit the orbit, many orbital integrations are required, so computing
the tilt between the stream and the orbit using an N-body simulation
on each integration would be extremely cumbersome.  
We need a simple method to infer the position and
velocity of the stream, given the orbit of its progenitor.  Our model
builds on the ideas presented in \citet{johnston98}.  That paper shows
the azimuthal phase at apocenter does not vary strongly with orbital
energy, i.e., along the stream.  From this observation, it infers that
the azimuthal phases of the different stream orbits are self-similar.
That is, the azimuthal phase $\Psi$ of a given orbit in the stream as
a function of time $t$ past pericenter is simply $\Psi(t) = \Psi_0 +
\Psi_{cen}(t/\tau + t_0)$, where $\Psi_{cen}$ is the phase of the
central orbit, the time scaling factor $\tau$ depends on the orbital
energy, and $t_0$ and $\Psi_0$ set the time and azimuth at the point
where the progenitor is disrupted, creating the stream.

We extend these ideas by suggesting that the stream stars with different
energies approximately follow a series of geometrically similar
orbits, differing only in their time {\it and length} scales.  The
orbits of the stream particles predominantly sample radii in the range
$5 < r < 150 \kpc$.  Over this range, our best-fitting model potential for M31 
is fairly close to a power law: $\Phi \propto r^k$ with index $k=-0.4 \pm
0.2$, significantly steeper than an isothermal halo.  Then we can use
the concept of mechanical similarity \citep{landau} to relate
different orbits that are similar in shape.  
Let the orbital time scale of such an orbit be proportional to
a time scaling factor $\tau$, with $\tau = 1$ for the orbit 
of the progenitor.  Then the
radius scales as $\tau^{2/(2-k)}$, the velocity scales as
$\tau^{k/(2-k)}$, and the energy scales as $\tau^{2k/(2-k)}$, while the
kinetic energy $T$ and total energy $E$ relate as $T = k \, E /(k+2)$.
To derive the phase space position of the stream, we consider a
sequence of values of $\tau$.  Let us choose the zero-point of time
$t$ to be the pericentric passage where the stars are disrupted.
(Our model explicitly assumes that the disruption is due to 
passage near M31, as opposed to some other cause such as an 
interaction with another satellite galaxy.)
Given the orbital position $\vec{r}_p(t)$ and velocity $\vec{v}_p(t)$ 
of the progenitor, we find the
position of a stream star with orbital scale factor $\tau$ as
$\vec{r} = \tau^{2/(2-k)} \vec{r}_p(t/\tau)$,
and the velocity as
$\vec{v} = \tau^{k/(2-k)} \vec{v}_p(t/\tau)$.  
The tilt of the stream
in phase space is then just a function of the time past pericenter.
(If several streams exist from successive pericentric passages, 
the temporal zero-point is different for each,
implying a different scaling factor $\tau$ for each at any
given azimuth as well.)  The
extent of the stream, in contrast, depends on the {\em spread} in
$\tau$ or equivalently in orbital energy.  This spread depends on the
satellite mass as well as its orbit, so we postpone detailed
consideration of the stream extent to the next section.  With our
model, the different orbits do not originate from precisely the same
location, but the difference in starting location is smaller than the
size of the progenitor itself so this is not a significant problem.

To check this approximate model, we evolve an N-body model of the
stream.  We assume that the progenitor of the stream is a dwarf galaxy
with total mass $2 \times 10^8 \msun$, consistent with the mass
estimates from the dynamical properties of the stream of
\citet{font04}.  We represent the satellite by a Plummer model, and
take its scale radius to be 0.3 kpc (note the half-mass radius of a
Plummer model is $\approx 1.3$ scale radii).

We use $N=32768$ particles in the satellite, with spline softening
length $\epsilon= 30 \, \mbox{pc}$; only about 0.1\% of the mass is
contained within one softening length of the center, indicating the
satellite is well resolved.  We initialize the particle distribution
with the {\sc ZENO} package of J. Barnes; the particle velocities are set by
solving an Abel integral for the energy distribution (see
\citealp{binney87}), giving an initial configuration very close to
equilibrium.  We evolved this satellite in isolation for 2 Gyr, and
found only negligible changes in the structure of the satellite.

\begin{figure*}
\includegraphics[width=16cm,bb=0 8 504 495]{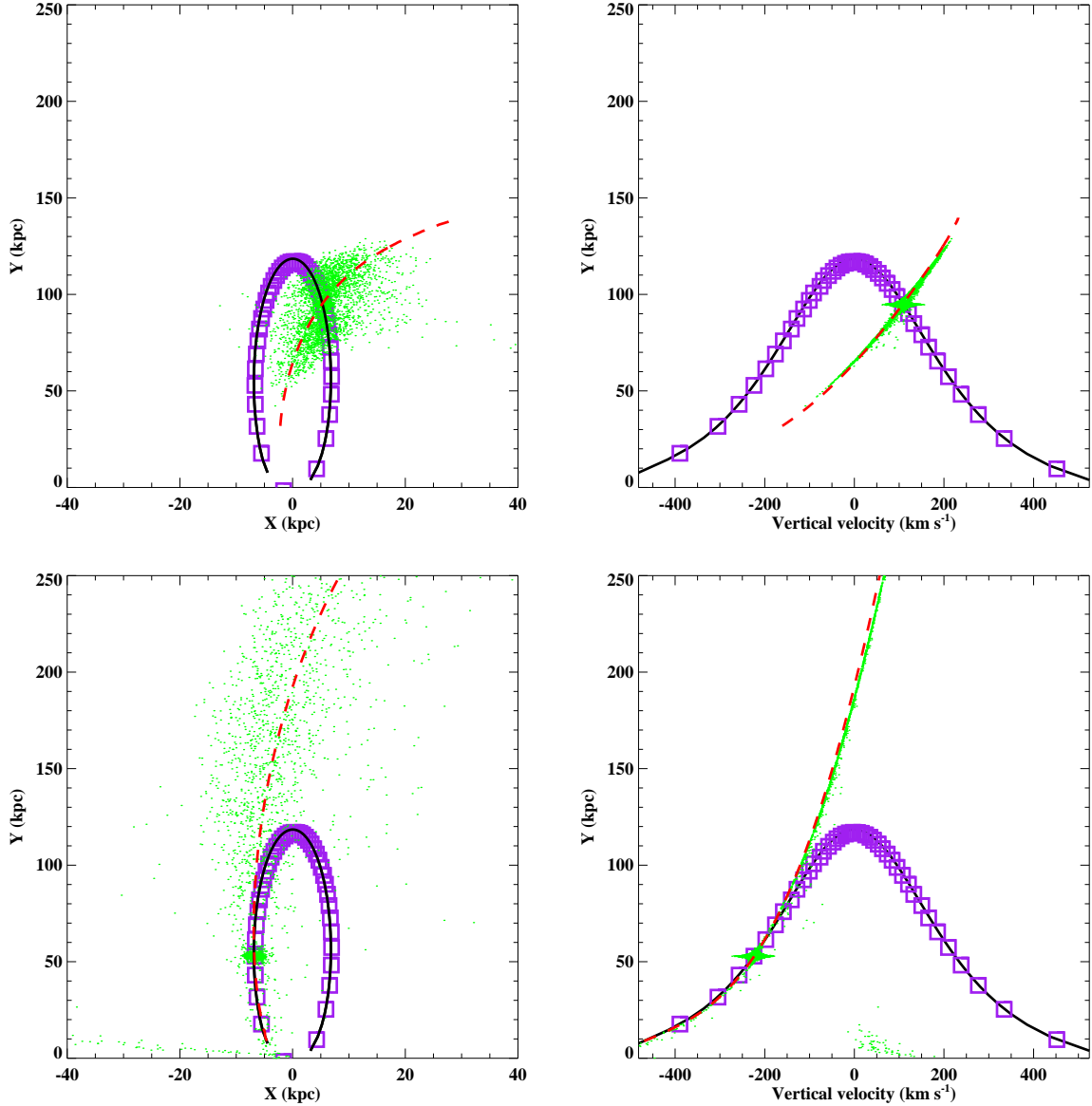}
\caption{
\label{fig.streamapprox}
Test of the accuracy of our approximation for the position of the stream
in phase space.  In these panels, green dots show the particles
in our N-body simulation of \S~2.2, which uses our best fit to the
southern stream data {\it neglecting} the stream-orbit tilt.
The solid line shows the orbit calculated in this potential.
Purple squares show the mean trajectory of the central 
particles of the satellite in this simulation, which closely follows
the orbit.  Red dashed lines show our approximation for the position 
of the stream.  We choose the coordinate system for this figure so 
that the satellite orbit executes a radial loop within the $X$-$Y$ plane,
which thus also contains the stream.
(Upper left): Spatial position of the particles, on an outbound 
portion of the orbit past the first pericentric passage.
The axis ratio has been distorted for readability.
(Upper right): Same as upper left, 
except shows the $Y$ velocity on the horizontal axis.
(Lower left): Same as upper left, except the satellite is now inbound.
(Lower right): Same as lower left, 
except shows the $Y$ velocity on the horizontal axis.
}
\end{figure*}
We place the satellite on an inbound orbit on the near side of M31.
We take this to be the first approach of the satellite.  In reality,
unless the progenitor orbit consists of an extremely lucky shot,
dynamical friction must be invoked to bring the satellite in to this
point.

We perform the calculation with the versatile parallel tree N-body
code {\sc PKDGRAV} \citep{stadel01,wadsley04}.  Our runs use the same fixed
Hernquist bulge, flattened exponential disk, and Navarro-Frenk-White (NFW) 
halo potentials as in our orbit integration code above.  The gravity tree 
uses an opening angle $\theta=0.8$ and the
node forces are expanded to hexadecapole order, while
the individual particle timesteps are limited by the acceleration criterion 
$\Delta \! t < \eta (\epsilon / a)^{1/2}$ with $\eta = 0.2$.

In Fig.~\ref{fig.streamapprox}, we demonstrate the accuracy of our
approximation for the stream.  
To display the results more clearly, we choose a coordinate system
just for this plot such that the radial loop after the first
pericenter lies in the $X$-$Y$ plane.
The plots show the trajectory of a single test
particle, and the median trajectory of the initially most-bound
particles in the satellite.  These nearly overlay each other, as
expected; the small difference is due to the dynamical effect of the
stripped outer portions on the inner core, as we can verify by
decreasing the satellite mass.  The upper and lower panels show the
positions and velocities of the stream particles at two different time
stages.  At early times, there is a significant tilt of the stream in
space relative to the orbit.  Later on, this tilt is lessened, but the
stream extends further in radius than the calculated orbit.  This is a
product of the systematic gradient in orbital energy along the stream,
where the trailing particles have higher energy \citep{johnston01}.
There is also a pronounced tilt in the space of $Y$ versus $v_Y$.
This relation is extremely tight, since the height reached
depends almost entirely on the orbital energy.  Phrased another way,
the large extent in physical space must be balanced by a small extent
in velocity space in order to conserve phase space volume.

Our approximation to the stream's phase-space location is shown as a
dashed line.  From here on we assume a potential slope of $k=-0.4$;
this is consistent with our power-law fit to the potential, and seems
to produce the best results on this plot as well.  Note the stream
approximation describes only the central ridge of the stream in phase
space, and predicts neither its extent along this ridge nor its width.
The stream approximation seems to be fairly accurate.  Some deviations
are apparent, but in general these are fairly small compared to the
width of the stream.  The approximation is certainly much better than
assuming the stream follows the orbit.  In the rest of the N-body
simulations in this paper, we generally find good agreement with our
stream approximation as well.  The agreement worsens slightly as the
satellite mass increases, since the self-gravity becomes more
important, but it still appears acceptable given the observational
errors.  The approximation thus satisfies our need for an easy way to
model the stream in a fitting routine, avoiding the need for many
N-body simulations.
\subsection{Fits with stream-orbit tilt}
We now repeat the fitting exercise above, incorporating the
stream-orbit tilt.  This adds an extra parameter, the current phase of
the progenitor along the orbit.  We express this as a fraction of the
{\em radial period}, i.e., $F_p \equiv t / t_r^{(P)}$, where $t$ is the time
since the progenitor went through the stream's last pericentric
passage, and $t_r^{(P)}$ is the radial period of the orbit of the progenitor.
(This is not to be confused with the {\em azimuthal} phase $\Psi$
employed by \citealp{johnston98} among others, though the two are
obviously related.)  
For a stream star, the orbital phase is instead $F = t / t_r$,
with $t_r$ the radial period of the particular star.  
The time scaling factor is then conveniently given by 
$\tau = t_r / t_r^{(P)} = F_p / F$.  We keep
$y(0)$ fixed, so we now have eight free parameters: six orbital
parameters, plus the halo parameter $f_h$ and the systematic shift
$\Delta \! z$ in the depth.  We test values for $F_p$ ranging from 0.5
to 2.5, or in other words from the apocenter of the stream's radial
loop to the one that is two radial periods ahead.  We find $F_p$ is nearly
unconstrained by the fit; the changes it introduces in the
stream location can be compensated for by changes in the other free
parameters.  We hence treat it as an external parameter, instead of
trying to fit it.  For the sake of discussion, we choose $F_p = 1.0$
in this section unless otherwise specified.

When we allow $f_h$ to vary freely, {\em without} the inclusion of the
galaxy fit term in $\chi^2$, we find best fit values of $f_h \gtrsim
4$ or $r_h \gtrsim 40$.  This is not only strongly disfavored by our
galaxy fits but implies a halo concentration $C_{200} \lesssim 7$,
which is at the bounds of physical plausibility (see Paper I).
This $f_h$ value is higher than that in \S~2.2, because the
gradient of orbital energy along the stream adds to the required
potential gradient.  However, once we add the term from the galaxy
fit, we find a best fit of $f_h \approx 1.2 \pm 0.4$ practically independent
of $F_p$, similar to the results in \S~2.2.  This implies a mass
$M_{125} = (7.4 \pm 1.2) \times 10^{11} \msun$.  The best fit orbital 
parameters for several choices of $F_p$ are listed in 
Table~\ref{table.streamfits} (orbits 2, 3 and 4).  
The reason for the lowering of $f_h$ from $\gtrsim 4$ to $\approx 1.2$
is that the variation in galactic $\chi^2$
dominates the variation from the stream.  This is partly because the
fit has another way to increase the computed potential gradient
besides changing $f_h$: $\Delta \! z$ is set to a lower value, placing
the near end of the stream at a smaller distance from M31.  
Recall that the systematic error in the depth is 16 kpc, compared to
the error in the transverse position of about 2 kpc. We are roughly looking
down the stream, so the distance uncertainty allows a large shift in
the radius and thus in the potential.

Even though the best-fitting $f_h$ is about the same as in \S~2.2, the
orbit is only about half as long as before; the radial period is
0.80 Gyr, apocenter is at 63 kpc, and pericenter is at 2.7 kpc.  The
stream-orbit tilt helps to explain why the stream is so strongly
radial; the initial conditions required to produce large
apocenter-pericenter ratios are rather less special than for an orbit
that follows the stream.  The apocenter and radial period decrease with $F_p$,
since it controls the relative sizes of the stream and the progenitor
orbit through the scaling factor $\tau$.  For example, with $F_p = 1.5$,
the best-fitting orbit has radial period 0.56 Gyr, apocenter 46 kpc, and 
pericenter 2.4 kpc.  The period varies only
slightly with $f_h$.  Typically, the projection of
the second lobe on the sky lies at a smaller position angle, or
northward of the NE side rather than eastward as in \S~2.2, but the
direction is sensitive to exactly where the orbit passes through the
disk.

\begin{figure*}
\includegraphics[width=16cm,bb=0 8 504 495]{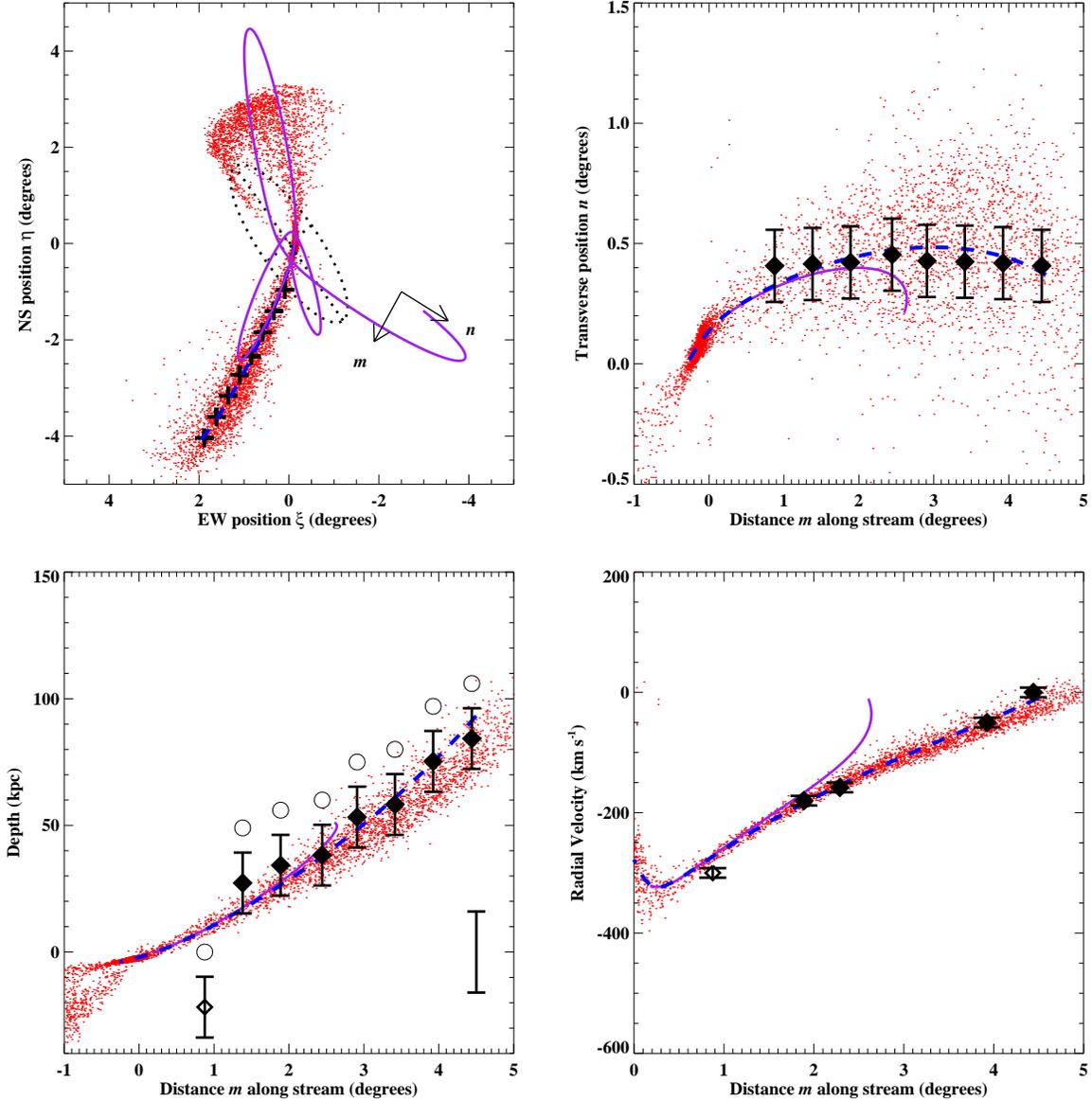}
\caption{
\label{fig.tiltorbit}
Results accounting for a tilt between the stream and the orbit.
Quantities shown in the panels are the same as in Fig.~\ref{fig.orbit}.
The orbit is shown with purple lines.  The corresponding
position of the stream calculated using our analytic approximation is
shown with blue dashed lines.  We conduct an N-body simulation using
initial conditions taken from this orbit; particles in this simulation
are shown with red points.  Observational results are shown with black symbols.
The remnant of the progenitor is a small clump near the center of M31,
in keeping with the orbital phase of $F_p = 1.0$.}
\end{figure*}
Using our best-fitting orbital parameters (with $F_p=1.0$), we now repeat our 
N-body simulation.  In this case we doubled the satellite mass to $4 \times
10^8 \msun$ and raised the satellite scale radius to $0.6 \kpc$, to get a
better match to the length of the stream, and set the softening
length to 0.1 of the new scale radius.  We again start the satellite
inbound towards the last pericentric passage of the
stream; the initial conditions for this point are given
as ``2e'' in Table~\ref{table.streamfits}.
Fig.~\ref{fig.tiltorbit} compares our orbit and the
resulting stream to the observations.  The agreement of the
observations with the stream approximation is excellent.  The
agreement with the N-body results is not quite as good, since as in
Fig.~\ref{fig.streamapprox} the approximation slightly
underestimates the azimuthal phase of the particles near pericenter,
but this offset is comparable to the observational errors.  As
expected, the progenitor has a more compact orbit than the stream
stars; the orbit turns around at about $2.5^\circ$, even though the
stream itself extends out to $4.5^\circ$.  With the chosen $F_p =
1.0$, the satellite lies against the M31 disk where it may have
escaped detection.

The stream's continuation on the opposite side of M31 is beginning to
develop a shell feature in Fig.~\ref{fig.tiltorbit}.  This feature
is not always apparent; it depends on the orbit orientation and
orbital phase.  The continuation of the stream is also significantly
rotated to the north compared to Fig.~\ref{fig.orbit}.  After the
first pericentric passage, the tidally stripped debris gradually
spreads out along the orbit, forming a well-defined stream.  After the
second passage, the stream acquires a significant width as well.  This
supports the idea that the narrow southern stream was indeed disrupted
less than one radial period ago.  The debris forward of the satellite
is significantly displaced from the position of the orbit, just like
the trailing debris.  This effect should be taken into account when
trying to match up possible stream orbits with the faint features in
the M31 disk.  
\section{Stream and Progenitor Properties}
In the previous section, we have considered the properties of the
stream that have a direct bearing on the orbital parameters.  We have
obtained estimates of the orbit of the progenitor, but the orbital
phase of the progenitor is unconstrained by the fit and introduces an
unwelcome degree of uncertainty to the orbit itself.  We also found
that the position of the next orbital loop could point in a variety of
directions, depending sensitively on the orbital parameters.  So
before we try to match the orbit to various observed features, let us
consider what we can learn about the orbital phase and other
properties of the progenitor from stream properties such as its
length, width, velocity dispersion, and luminosity.  These results
should stand up regardless of the identification of the progenitor.

\begin{table*}
\begin{minipage}{155mm}
\caption{Parameters of the N-body simulations used to estimate the 
stream quantities.  Each run is labeled by an arbitrary integer.
The orbital phase of the progenitor is denoted by $F_p$, and the 
initial conditions corresponding to the orbit label are listed in 
Table~\ref{table.streamfits}.  
$M_s$ is the total initial satellite mass, and
$a_s$ is the scale length of the Plummer model.  In each case we take
the softening length to be $0.1 a_s$.  
The stream mass $M_{stream}$ is defined as the total mass of N-body
particles within the phase space region outlined by
Equations~\ref{eqn.mboundary}--\ref{eqn.mboundary.end}.  
The fraction of stars in the southern stream still moving 
radially outward is given by $f_{\it outbound}$.  
The values
for the angular FWHM of the stream are derived from
Gaussian fits with variable baselines; the value marked with a
question mark differs significantly from the true width of the stream,
owing to the mismatch between the narrow field and the large true width.
Finally, $\sigma_{vr}$ gives the 
rms dispersion in the velocity relative to the observer of the
stars in the stream region, computed by binning the particles 
as in Fig.~\ref{fig.streamvdisp} and subtracting the mean
velocity within the bin before finding the total dispersion.}
\label{table.runs}
\begin{center}
\begin{tabular}{llllllllc}
\hline\\  
 Run &  $F_p$ &  $M_s$  &  $a_s$  &  Orbit  
  &  $M_{stream}$ & $f_{\it outbound}$  &  FWHM  &  $\sigma_{vr}$  \\
     &        &$(10^8 \msun)$   &  (kpc) &      
  & $(10^8 \msun)$&                 &  (${}^\circ$)  & $(\mbox{km}\,\mbox{s}^{-1})$   \\
\hline
  13  &  1.0 &  0.5  &  0.3    &  2e  &  0.15  & 0.000 &  0.26  &  11   \\
  11  &  1.0 &  4.0  &  0.6    &  2e  &  1.0   & 0.021 &  0.53  &  14   \\
  14  &  1.0 &  10   &  0.815  &  2e  &  1.9   & 0.067 &  0.69  &  17   \\
  12  &  1.0 &  32   &  1.2    &  2e  &  3.5   & 0.149 &  0.91  &  25   \\
  23  &  1.5 &  0.5  &  0.3    &  3e  &  0.0003& 0.000 &  0.11  &   5   \\
  21  &  1.5 &  4.0  &  0.6    &  3e  &  0.33  & 0.000 &  0.32  &  12   \\
  24  &  1.5 &  10   &  0.815  &  3e  &  1.6   & 0.008 &  0.55  &  16   \\
  22  &  1.5 &  32   &  1.2    &  3e  &  4.1   & 0.078 &  0.39? &  24   \\
  34  &  2.0 &  10   &  0.815  &  4e  &  0.38  & 0.003 &  0.34  &  14   \\
  32  &  2.0 &  32   &  1.2    &  4e  &  3.2   & 0.030 &  0.86  &  22   \\
\hline
\end{tabular}
\end{center}
\end{minipage}
\end{table*}

We begin by generating an extended sample of N-body simulations.  We
use three different orbits, namely 2e, 3e, and 4e in
Table~\ref{table.streamfits} which are the best fits for $F_p = 1.0$,
1.5, and 2.0 respectively.  We again select a point on the inbound
loop before pericenter as a starting point for each simulation.  For
each orbit, we test several values of the satellite mass.  We adjust the
scale length of the satellite along with its mass to keep the
characteristic density constant, ensuring similar amounts of satellite
disruption in each run.  The relation between the mass and velocity
dispersion in these runs is consistent with that observed in dwarf
galaxies (e.g., \citealp{derijcke05}), suggesting this characteristic
density is reasonable.  In all other respects, the simulations are
like those described before.  The runs and some statistics derived
from them are listed in Table~\ref{table.runs}.  The run already
displayed in Fig.~\ref{fig.tiltorbit} is ``run11'' in this table.

We examine the resulting particle distributions at the timestep
determined by $F_p$.  One trend is obvious, from the positions of the
particles in phase space: as the progenitor mass increases at a fixed
orbit, the length, width, and velocity dispersion of the stream
increase significantly.  This suggests we can obtain an estimate of
the progenitor mass by considering these properties of the stream.
This work follows up the ideas in \citet{font04}, but we improve the
results by using our revised orbits and some modified formulae, and
checking the results with our N-body simulations.
\subsection{Length and stellar mass of the stream}
The length of the stream is related directly to the spread of energies
of its stars.  For a satellite of
mass $M_s$ and pericenter $R_s$ in an isothermal halo,
\citet{johnston98} found the energy range of tidal debris tends to be
spread out over the range $-2 \epsilon < \Delta E < 2 \epsilon$, where
$\epsilon = V_c^2 (G M_s / V_c^2 R_{peri})^{1/3}$, $M_s$ is the
satellite mass, $R_{peri}$ is the pericentric distance, and $V_c$ is
the circular velocity at pericenter.  By examining our small sample of
N-body simulations, we find results consistent with the mass scaling
in this formula.  However, we find a somewhat smaller spread: in our runs, 
97\% of the particles have $\Delta E \lesssim 1.3 \epsilon$ after the first 
close pass.   ($\Delta E$ is distributed roughly symmetrically around 0,
but we use a one-sided test here since in some simulation outputs the forward 
stream is affected by a second pass through pericenter.)  Aside from this 
change, the derivation of \citet{font04} for the length of the stream may 
not be valid for large energy spreads or for non-isothermal halos.  We
hence derive the relation as follows.  

\begin{figure}
\includegraphics[width=84mm,bb=0 8 504 485,]{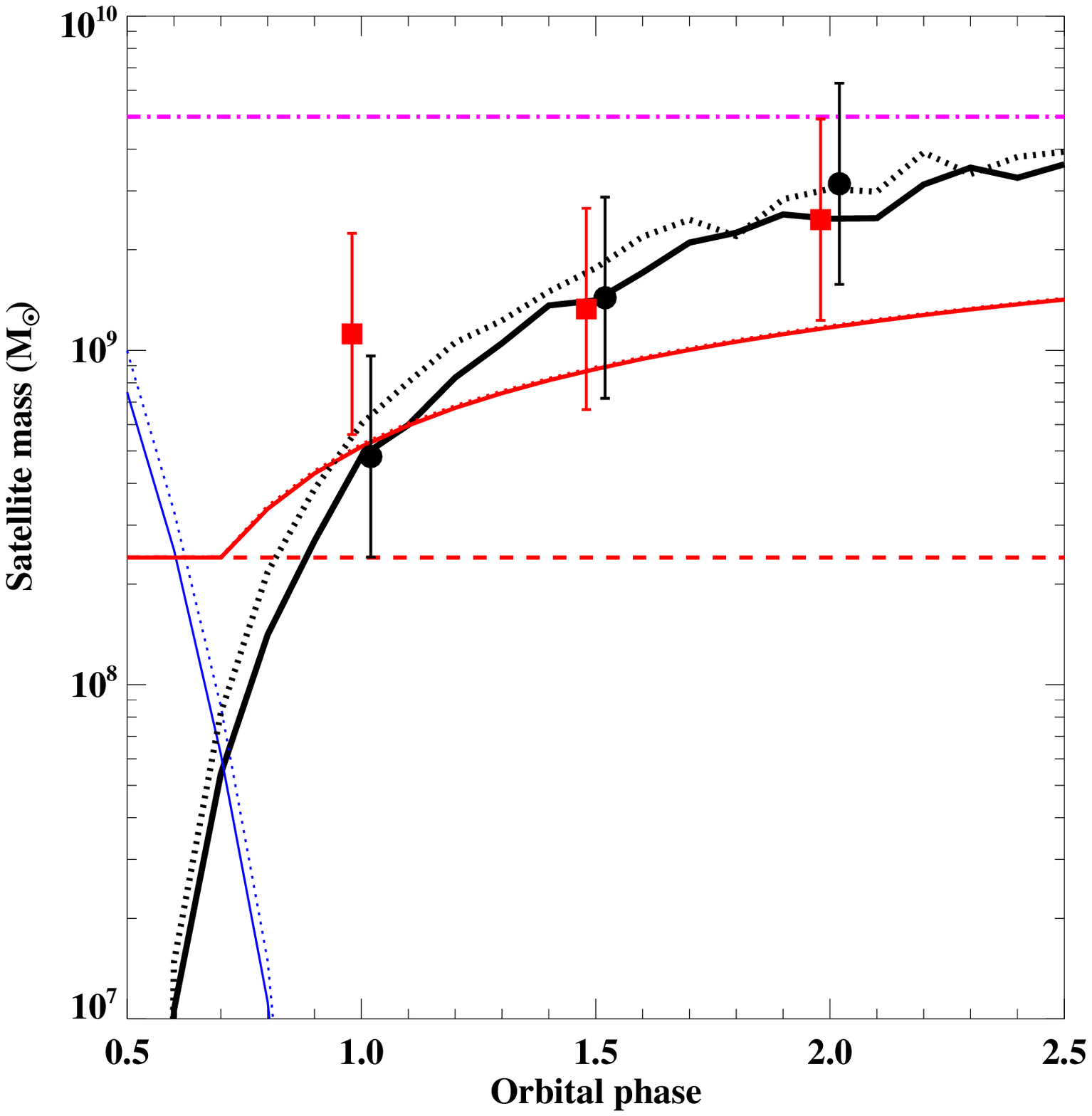}
\caption{
\label{fig.satmass}
Shows estimates of the required mass $M_s$ of the progenitor satellite
as a function of orbital phase $F_p$.  The solid and dotted 
rising black lines show the
satellite mass needed to make the stream extend to apocenter, derived
from Equation~\ref{eqn.satmass}.  These lines use the best fit orbital
parameters for the two values of $f_h = 1.2$ (solid) and 2.0 (dotted).
The noise at high orbital phase corresponds to jiggles in the
parameters from our $\chi^2$ fit, which are not very stable in this
region.  The black circles show the estimate of the mass from our
suite of N-body simulations, obtained by interpolating the fraction of
stars that are still moving outwards at the timestep corresponding to
orbital phase $F_p$.  The blue, rapidly declining lines at the left
side of the figure show the {\em minimum} satellite mass needed to
make the stream extend to its closest observed point (Field 8), again
using the two cases $f_h = 1.2$ (solid) and 2.0 (dotted).  The horizontal
red dashed line shows the observed stellar mass \protect\citep{ibata01}.
The red gently rising solid line shows the estimate of the progenitor mass 
needed to put this amount of stellar mass in the stream, as derived from
Equation~\ref{eqn.satlum}.  The red squares show the corresponding
estimate of the mass from our suite of N-body simulations, obtained by
interpolating the mass of particles found in the stream.  The two data
points at each orbital phase are slightly offset from each other for
the purpose of clarity.  Finally, the magenta dot-dashed line shows
the estimate of the progenitor's mass from the metallicity of the
stream stars \protect\citep{font04}.}
\end{figure}
Our mechanical similarity
argument in \S~2.2 implies that the orbital energy $E = E_0 + \Delta
E$ depends on the time scaling factor $\tau$ as
$E/E_0 = 1 - \Delta E / |E_0| = \tau^{2k/(2-k)} \approx \tau^{-1/3}$.
(The energy $E_0$ of the progenitor is negative, if we are using
the approximate power-law form for the potential; hence
$\tau$ increases with $\Delta E$).  We then find
\begin{equation}
\label{eqn.satmass}
M_s = \frac{R_{peri} \, |E_0|^3}{1.3^3 G \, V_c^4} 
  \left( 1 - \tau_{max}^{-1/3} \right)^3 \; .
\end{equation}
Here $\tau_{max}$ is the value of the time scaling factor at the end
of the stream, and equals the ratio of the orbital phase of the
progenitor $F_p$ to that of the stream stars at stream end.  
In our case we know that the stream extends just about
to apocenter where the phase $F = 0.5$, so $\tau_{max} \approx 2 F_p$.  
We obtain
$R_{peri}$ and $E_0$ from the orbit of the progenitor, and we take
$V_c = 250 \kms$.  

We can also derive a constraint on the mass from the other end of the 
southern stream, Field 8 of \citet{mcconnachie03} which is the 
closest field to M31 and the furthest along in the orbit.
Since this field does not necessarily represent the end of the stream,
we can only derive a {\em minimum} satellite mass, valid when
the progenitor is assumed to lie within the visible extent of the
stream:
\begin{equation}
\label{eqn.satmasslimit}
M_s > \frac{R_{peri} \, |E_0|^3}{1.3^3 G \, V_c^4} 
  \left( \tau_{F8}^{-1/3} - 1 \right)^3 \; .
\end{equation}
We obtain $\tau_{F8}$ by matching the $m$ coordinate of Field 8 to the
calculated $m$ coordinate of the stream as a function of $\tau$.

We can now estimate the satellite mass required to produce the visible
extent of the stream.  We use the best fit parameters evaluated for a
series of $F_p$, restricting the value of $f_h$ to the values 1.2 (our
best fit for most values of $F_p$) or 2.0 ($\tsim 2\sigma$ higher).
The preferred mass from Equation~\ref{eqn.satmass} is shown as the black lines 
in Fig.~\ref{fig.satmass}.  The estimates from the two values of $f_h$
do not differ too much, indicating our result for the required
progenitor mass is fairly robust.  The main thing to observe is the
strong correlation between the orbital phase and the mass.  The
primary reason for this is the dependence on $F_p$ through
$\tau_{max}$.  As the orbital phase increases for a {\em fixed} orbit,
the value of the mass must increase in order to create a stream out to
apocenter.  There are also changes in the orbital quantities $E_0$ and
$R_{peri}$ as the best-fitting orbit changes as a function of $F_p$, but
these are smaller effects and nearly cancel out in any case.  The
minimum satellite mass from the forward extension in
Equation~\ref{eqn.satmasslimit} is also shown in the plot.  This
places only a weak limit on the mass and orbital phase, which will
shortly be superceded.

The estimate of the satellite mass from the stream length has
significant uncertainty, since we are relying on an
analytic scaling relation which has not been rigorously established.  
Also, the ``end'' of the stream is not very precisely defined; since
the energy distribution of stream stars drops off rapidly but continously 
with $\Delta E$, if the observationally apparent end of the stream extends
to apocenter, there should be a small fraction of stars ($\tsim 3\%$)
that are still 
moving outwards towards apocenter. We can test the formula with our N-body 
simulations. Table~\ref{table.runs} lists the fraction of the total stars 
moving outwards; this is clearly a very steep function of the progenitor mass
for a given orbital phase. If we require this fraction to be 3\% of the total 
stars in the progenitor, and interpolate the results in the table, we
obtain an estimate of the progenitor mass. (This mass is not very sensitive
to the adopted threshold, since the number of stars at large $\Delta
E$ is falling rapidly with $E$).  These estimates, shown as filled
circles in Fig.~\ref{fig.satmass}, seem to be consistent with
Equation~\ref{eqn.satmass}.  This is not too surprising, since we used
these runs to calibrate the numerical constant in this equation, but
it is reassuring nevertheless.

The stellar luminosity in the stream gives us an alternate means of
estimating the progenitor mass.  The red dashed line in 
Fig.~\ref{fig.satmass} shows the estimated stellar mass 
$M_{stream} \sim 2.4 \times 10^8 \msun$ of the stream \citep{ibata01}.  
Allowing for dark matter, this gives a lower limit on the
total mass. This immediately produces a tighter limit on the mass and
orbital phase than the length of the stream alone.
However, the fraction of stars that lie within the
stream region can be quite small, and varies strongly with the mass
and orbital phase.  For the sake of argument, let us assume that the
energy of the stream stars is distributed uniformly in the range $-1.3
\epsilon < \Delta \! E < 1.3 \epsilon$.  We can use this to derive the
fraction of stars that lie within the allowed range
$\tau_{F8} < \tau < \tau_{max}$.  In turn, this
raises the estimate of the mass of the progenitor to
\begin{equation}
\label{eqn.satlum}
M_s =  2 M_{stream} \frac{1 - \tau_{max}^{-1/3} }
{ \tau_{F8}^{-1/3} - \tau_{max}^{-1/3}} \; .
\end{equation}
As long as the stream overlaps the observed length, out to Field 8
(where the steeply rising black and falling blue curves cross), 
this equation gives more
restrictive limits on the progenitor mass than the observed stellar
mass alone.  The results are shown as the solid red curve in
Equation~\ref{fig.satmass}.  This curve is a {\it lower limit} on the
total satellite mass, since if the progenitor is partly composed of
dark matter, or if it is not fully disrupted, an even larger mass
is needed to match the observed stream luminosity.  Hence,
in Fig.~\ref{fig.satmass}, $M_s$ and $F_p$ should
lie on the portion of the black curve that lies above the red curve.
This implies $F_p \gtrsim 0.9$ and $M_s \gtrsim 4 \times 10^8 \msun$.  
The square points in the figure are obtained by interpolating the 
progenitor mass $M_s$ as a function of the mass $M_{stream}$ of stars in the 
stream region; this is defined below in 
Equations~\ref{eqn.mboundary}--\ref{eqn.mboundary.end}) and given
in Table~\ref{table.runs}. The square points lie slightly above the 
curve, in part because the progenitor is not quite fully disrupted,
and also because the stars are starting to leak out from the sides of 
our stream region as well as from the ends.  This validates our argument
that the red curve is a lower limit to the mass.

The results in this plot are not extremely precise, but still several
points are evident.  First, the progenitor most likely lies ahead of
the stream in its orbit.  Second, the required mass is strongly
correlated with the orbital phase.  Third, the lower limit on the mass
from the stream luminosity roughly matches the estimate of the mass
from the stream length.  This suggests that there was not much dark
matter before last pericenter, and the progenitor was largely disrupted
at last pericenter.  It is quite plausible that the outer
dark-matter-dominated regions were stripped by tides before the
stellar core was disrupted.  Fourth, the coincidence of these two
curves is only obtained if we assume the stream is young (stripped at
last pericenter).  If we assume instead that the stream resulted from
two pericentric passages ago, the estimated mass from the stream
length (black curve) shifts down by a large factor, but the lower limit
on the mass from the stream's luminosity (red curve) does not change
much.  In other words, the total mass implied by the directly observed 
mass in stream stars would be enough to spread out such an old stream
to an unacceptable degree.  Unless our estimate for the orbit is badly
in error, we can rule out this model.

Using the stream's metallicity, \citet{font04} estimated the
progenitor had a stellar mass of $M_s = 5 \times 10^9 \msun$.  This
estimate has an uncertainty of perhaps a factor 3, given the scatter
in the metallicity-luminosity relation.  Thus it is consistent with
our similarly rough estimates of the satellite mass for almost any
orbital phase, but the agreement is best for $F_p \sim 2.5$.
Alternatively, the progenitor may have lost stars to gravitational
tides earlier in its life, moving its mass $M_s$ at last pericenter off the
mass-metallicity relation.  In this case we expect $F_p \lesssim 2.5$.
\subsection{Other stream measures}
We now consider other observational measures of the stream compared to
our N-body simulations.  Our set of N-body models shows particles at a
wide range of transverse distances from the center of the stream, not all of
which would be picked up in the current observational surveys.  The
extent of this satellite debris depends on the mass and radius of the
satellite as well as the pericenter distance $R_p$.  To exclude the more
extreme stellar trajectories in the stream, we define a ``stream
star'' as one lying in the region
\begin{eqnarray}
0.60^\circ & < & m < 4.70^\circ \label{eqn.mboundary} \\
-0.14^\circ & < & n < 0.62^\circ\\
-30 \kpc & < & d' - 20 m - \Delta \! z < 60 \kpc \\
-410 \kms & < & v'_r -70 m < -210 \kms  \; . \label{eqn.mboundary.end}
\end{eqnarray}
The length and transverse width of this boundary approximately match
the fields in \citet{mcconnachie03}.  The $\Delta \!  z$ term
compensates for the fitted offset between our distance
calibration and that of \citet{mcconnachie03}.  We chose a more
generous selection boundary in velocity space than used by
\citet{ibata04} and \citet{raja05}, because we do not have the problem
of excluding a background density of stars; in any event the
velocity-position relationship is tight enough that this is not a
major factor.

\begin{figure}
\includegraphics[width=84mm,bb=0 8 504 485]{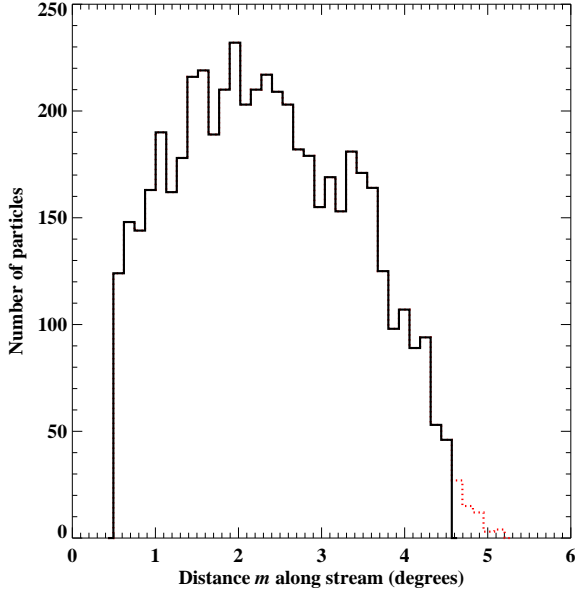}
\caption{
\label{fig.streamlength}
Shows the distribution of N-body particles along the stream, using run
24.  Position $m$ along the stream is defined the same way as in
Fig.~\ref{fig.tiltorbit} (so its sign is reversed from Fig.~4 of
\protect\citealp{mcconnachie03}).  Cuts are made to the sample as
described in the text.  The dotted line continues the histogram
without the sample cut in Equation~\ref{eqn.mboundary}.}
\end{figure}
\begin{figure}
\includegraphics[width=84mm,bb=0 8 504 485]{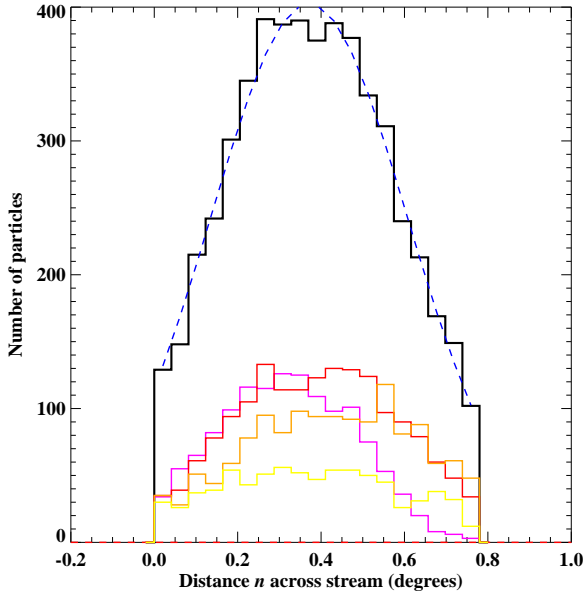}
\caption{
\label{fig.streamwidth}
Shows the distribution of N-body particles as a function of $n$,
the coordinate direction
perpendicular to the stream, using run 24.  Position $n$ across the
stream is defined the same way as in Fig.~\ref{fig.tiltorbit} (so
the sign is reversed compared to fig.~3 of \protect\citealp{mcconnachie03}).
The NE side of the stream is to the left.  Cuts are made to the sample
as described in the text.  
The low-amplitude magenta, red, orange, and yellow histograms 
(highest to lowest at the peak) show the distribution
in quadrants of increasing distance along the stream,
while the black histogram shows the total.
The blue dashed curve shows a Gaussian fit with a variable
baseline, having FWHM $0.57^\circ$.}
\end{figure}
Fig.~\ref{fig.streamlength} shows the distribution of N-body
particles along the stream.  This is analogous to fig.~4 of
\citet{mcconnachie03}.  In comparing the two, we should remember that
``background'' (non-stream) stars probably contribute a substantial
and spatially varying baseline to the observed histogram.  By
monitoring the behavior of the histogram for different timesteps and
different runs, we see that this plot is essentially another way to
check the length of the stream.  If the stream is too long, the
histogram will rise towards its end; whereas if the stream is too
short, it will cut off in the middle of its range.  To the limited
extent we can tell, given the background issue, the distributions
appear to agree quite well.  This again indicates our progenitor mass
and pericenter are reasonable.

In Fig.~\ref{fig.streamwidth}, we display the distribution of N-body
particles in the direction normal to the stream.  This can be compared
to fig.~3 of \citet{mcconnachie03}, though again that figure
contains background stars not present in our run.  The stream in our
run happens to be centered in our chosen selection region (the
observed stream is slightly offset from the selection region used
here). We find the rise in the number on the NE side is comparable to
that in the observed stream (the other side is cut off by the field
boundary in the observations).  However, some outlier particles are
present in non-Gaussian tails outside the width of the fields, as can
be seen in the top panels of Fig.~\ref{fig.tiltorbit}.  Stars
showing this behavior would probably fade into the general halo
population in surveys using number counts alone.  In fact, it is
possible that they make some contribution to the wide-field,
minor-axis photometric surveys of \citet{durrell01,durrell04}.  These
fields are far enough away from the stream that, in our ``run24'',
$\lesssim 5 \times 10^5 \msun$ of stream material is covered by the
two fields in this survey.  From the relative surface brightnesses in
the stream and in the survey fields, we estimate this represents only
a few percent of the total mass of halo stars in these
fields. However, this contamination would be very sensitive to the
orbit, size, and radial mass profile of the incoming progenitor, so we
cannot rule out the possibility that it contributes significantly to
the observed population in these surveys.  Interestingly, these
surveys found a metallicity distribution peaking at $\mbox{[m/H]} \sim
-0.5$, similar to the metallicity of the stream.  In general, these
outlier stars will share in the distance and velocity signature of the
stream, so they may be recognizable in spectroscopic surveys.

\citet{font04} suggested that the spatial width of the stream (defined
to contain 80\% of the stream luminosity, which for a Gaussian 
is the FWHM) at a physical distance $R$
from M31 could be expressed as $w = s R$, where $s \equiv (G M_s / V_c^2
R_{peri})^{1/3}$.  Using this estimate, they fitted a Gaussian 
with a floating baseline to the transverse distribution of the stream
stars and inferred $M_s \approx 1.3 \times 10^8 \msun$; this is quite low
compared to their estimate from the stellar metallicity.  Font \ea
dismissed this discrepancy, noting that if the progenitor was
partially supported by rotation it might reduce the width of the
stream.  

We mimic their procedure by fitting a Gaussian with an arbitrary
baseline to the N-body runs.  The resulting estimate for the FWHM is
given in Table~\ref{table.runs}, which shows that it increases
systematically with $M_s$.  We find that this is not a completely
robust procedure to recover the true width of the stream, because the
window is narrow and the non-Gaussian tails can masquerade as a
baseline.  Nevertheless, in most cases we recover nearly the same
answer as we would with twice as wide a window, so their width
measurement is probably accurate.  Font et al.'s analytic formula
systematically overpredicts the estimated width in our runs.  However,
if we multiply their formula for $w$ by 0.4, we get fairly good
agreement overall.  It may be that their formula simply needs this
scaling constant, which raises the mass at a given width by more than
an order of magnitude, although more thorough testing of this is
warranted.  The best agreement of our N-body runs with the observed
width is obtained for $M_s \sim 10^9 \msun$, consistent with our other
estimates.  The mass estimated this way is not very precise but is
probably good to a factor 3.
If we interpolate the results in Table~\ref{table.runs}
to find the FWHM at a stream mass of $M_{stream} = 2 \times 10^8 \msun$,
as we did with the mass earlier, we find results of 
$0.71^\circ$, $0.50^\circ$, and $0.70^\circ$ 
for $F_p = 1.0$, 1.5, and 2.0 respectively.
Although the precision of these estimates is low, they indicate that
models that reproduce the stream {\it mass} also reproduce the
stream {\it width} fairly well, independent of the orbital phase.

\begin{figure}
\includegraphics[width=84mm,bb=0 8 504 485]{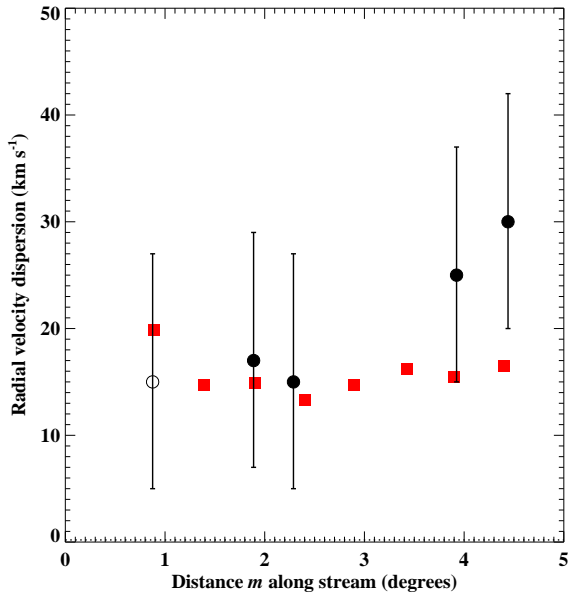}
\caption{
\label{fig.streamvdisp}
Shows the line-of-sight velocity dispersion of N-body particles in run
24 (red squares), as a function of position $m$ along the stream.  The
points are computed by taking particles within the stream region
(Equations~\ref{eqn.mboundary}--\ref{eqn.mboundary.end}),
binning them in $m$, and finding the standard deviation of the
velocity of the particles in each bin.  Black circles with error bars
show the observed velocity dispersion in five fields from
\protect\citet{raja05} and \protect\citet{ferguson04}. We indicate the
field in Ferguson \ea closest to M31 with an empty symbol due to its
potential contamination with the disk.  A rough error bar of $10 \kms$
is assumed for each observed point.}
\end{figure}
Fig.~\ref{fig.streamvdisp} shows the rms velocity dispersion at
several points along the stream.  Observed values of the velocity
dispersion are shown in the figure as well; these are quite uncertain,
given the small number of stars and the possibility of contamination
by additional structure in the background density.  The dispersion in
the N-body model is certainly consistent with the observations,
suggesting our progenitor mass is roughly correct.  

\citet{font04} suggested $M_s \gtrsim 2.0 \times 10^8 \msun$ from the
velocity dispersion; they noted that this is a lower limit since the
observed portions of the stream may be dynamically colder than the
progenitor itself, which we indeed find to be the case (cf.\ the
right-hand panels of Fig.~\ref{fig.streamapprox}).  We find the
central velocity dispersion of the progenitor is $\tsim 25 \kms$,
while the {\it M31-centric} radial velocity dispersion as a function
of {\it M31-centric} radius drops to very low values ($\lesssim 5
\kms$) throughout the stream.  The larger line-of-sight dispersion in
Fig.~\ref{fig.streamvdisp} is primarily due to the fact that the
stream stars form a fattened, tilted tube in three dimensions, and
thus a single line of sight probes particles at a range in M31-centric
radius which depends on the tube's thickness.  As a result, there is a
good correlation between the transverse width and the velocity
dispersion in Table~\ref{table.runs}.

Font et al.\ also suggested that the velocity dispersion, should
increase sharply near apocenter.  In contrast,
Fig.~\ref{fig.streamvdisp} shows it to be roughly constant to the
last measured point, at nearly the end of the stream.  The difference
may arise from Font et al.'s use of azimuthal binning rather than our
binning in transverse position $m$.  It is difficult to compare our
results directly, as the stream particles in our runs occupy only a
small range in azimuth, blurring any correlation that might exist.

If we again interpolate the results in Table~\ref{table.runs},
this time to find the velocity dispersion 
at a stream mass of $M_{stream} = 2 \times 10^8 \msun$,
we find results of  
18, 18, and $20 \kms$ for $F_p = 1.0$, 1.5, and 2.0 respectively.
Again, there does not seem to be a strong trend with orbital phase
for a {\it fixed} mass of stars in the stream.

In summary, we find that the various observed properties of the stream
give consistent indications of the mass.  Our simulation with a mass
of $M_s = 10^9 \msun$ and an orbital phase of $F_p = 1.5$ is the
one that is most consistent with the observed properties of the
stream.  There is some leeway in these quantities, however, as long as
the mass and orbital phase increase together in keeping with
Fig.~\ref{fig.satmass}.  Aside from detection of the progenitor
itself, radial velocity surveys will probably be
the most powerful tool to constrain the progenitor mass.  Radial
velocity surveys to a greater transverse extent may help to
differentiate between unrelated halo stars and the sideways extension
of the stream, which would give additional constraints on
the properties of the progenitor.
\section{Forward Continuation of the Stream}
From the analysis in the previous section, we now have a better idea
of the mass and orbital phase of the progenitor.  We now consider the
continuation of the stream past the next pericenter.  For the fit shown 
in Fig.~\ref{fig.tiltorbit}, the second lobe projects about $30^\circ$
counterclockwise of M31's NE major axis, 
but this direction is sensitive to the orbital parameters.
If we impose further
observational constraints, this loop can point in a variety of directions
which encompass a number of faint features observed near M31.
As photometric and velocity surveys go deeper and deeper, more faint
features will probably continue to turn up, and it is quite possible
that the progenitor is not even a currently identified feature.
Nevertheless, it is worth testing which of the currently identified
features can be the progenitor.

Table~\ref{table.prog} gives the observational data for a number of
these features in the disk and inner halo regions of M31.  We perform
fits in the same manner as in \S~2.4, but we add further data points
for the sky position of the progenitor, and its depth or radial
velocity if these are known.  Most of these possible progenitors
require a phase $F_p > 1.0$.  Once the progenitor passes through
pericenter for a second time, a second stream develops, which again
does not follow the orbit.  However, in our N-body simulations, the
progenitor itself follows the calculated orbit fairly closely, though
there are some slight deviations due to the self-gravity of the
satellite matter.  The position and velocity of the stream thus can be
derived directly from the calculated orbit and the orbital phase
$F_p$.  

The errors on the observational data are in many cases not well
specified, so we have chosen reasonable values in order to perform the
fit.  For example, some of the features are not very well-defined
spatially, including the Northern Spur, the Merrett et al.\ planetary
nebulae (PNe), and the Eastern Shelf.  We assume positional errors of
0.05--$0.2^\circ$, depending on how well-defined the object is.  In
Table~\ref{table.prog}, if the distance or velocity are not known for
a particular feature, we supply the results from our fit in brackets.
In some cases there may be alternate solutions to the one we have
computed, where the progenitor has a different orbital phase,
especially if only the sky coordinate of the feature is known.  We now
discuss the suitability of the following features as progenitors.

\begin{table*}
\begin{minipage}{150mm}
\caption{Kinematic data and predictions for possible progenitors of
the stream.  Brackets indicate that the field is a prediction from our 
best-fitting orbit, as opposed to observational data.  $d'$ and $v'_r$ 
give the radial displacement and velocity relative to M31.  The depth
offset $\Delta \! z$ and orbital phase $F_p$ are taken from our fit.
The final entry gives the value of reduced $\chi^2$ from our fit.  The
number of degrees of freedom in each case is 13 plus the number of
observational constraints in the table.  }
\label{table.prog}
\begin{tabular}{@{}llllcccc}
Field/Name & $\xi$~(deg) & $\eta$~(deg) & $d'$~(kpc) & $\Delta \! z$~(kpc) & 
$v'_{r}$~(km~s$^{-1}$)  & $F_p$  & $\chi^2 / N_{deg}$ \\
\hline
Linear continuation & $-1.0 \pm 0.2$ & $+1.3 \pm 0.2$ &  $-35 \pm 12$  &  $-13$  & [$-70$]  & 1.1 & 1.8 \\
NGC 205       	& $-0.5 \pm 0.05$ & $+0.4 \pm 0.05$ & $50 \pm 12$  &   ---   & $+55 \pm 10$    & --- & --- \\
M32           	& $+0.0 \pm 0.05$ & $-0.4 \pm 0.05$ &  $0 \pm 12$  &  $-15$  & $+100 \pm 10$   & 2.0 & 1.0 \\
Merrett NE PNe  & $+1.0 \pm 0.2$ & $+1.4 \pm 0.2$ & [$-40$] & $-6$    & $-190 \pm 10$   & 1.1 & 0.8 \\
Northern Spur 	& $+0.7 \pm 0.2$ & $+1.8 \pm 0.2$ & [$-21$] &  $-13$  & [$-127$] & 1.1 & 0.6 \\
Northern Spur, $v$ constraint	& $+0.7 \pm 0.2$ & $+1.8 \pm 0.2$ & [$-12$] &  $-13$  & $200 \pm 30$ & 1.8 & 1.3 \\
And NE       	& $+1.2 \pm 0.1$ & $+3.0 \pm 0.1$ &  [$-21$]  &  $-16$  & [14]     & 1.5 & 0.5 \\
Eastern Shelf  	& $+1.8 \pm 0.2$ & $+0.6 \pm 0.2$ & [$-56$] &     8   & [$-92$]  & 1.3 & 1.3 \\
And VIII      	& $+0.1 \pm 0.2$ & $-0.7 \pm 0.2$ & [$-53$] &    15   & $-205 \pm 20$   & 1.2 & 2.1 \\
\hline
\end{tabular}
\end{minipage}
\end{table*}
\begin{figure}
\includegraphics[width=84mm,bb=0 8 504 485]{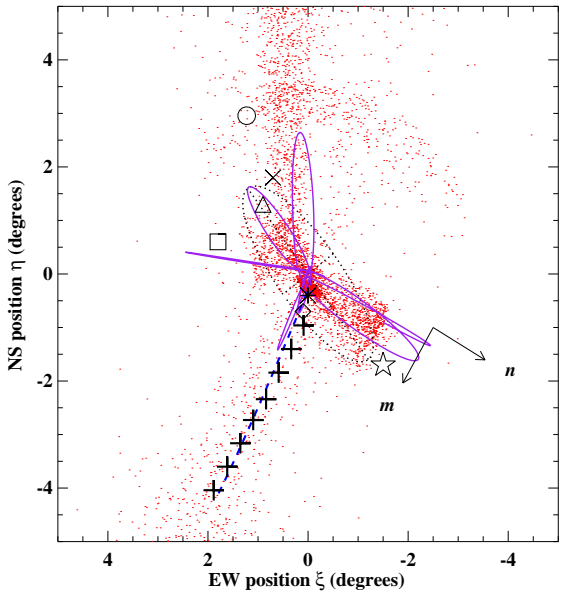}
\caption{
\label{fig.m32fit}
Best-fitting orbit passing through M32, and the particle distribution in
the corresponding N-body simulation.  Crosses, orbits, and dots have
the same meaning as in Fig.~\ref{fig.tiltorbit}.  The particles
corresponding to M32 itself are barely visible in this plot as the
clump just south of the center of M31.  The point density in the
stream is smaller than in Fig.~\ref{fig.tiltorbit}, because we use a
larger particle mass in this run to match the larger mass of the
progenitor.  Observed positions of some of the faint features are
plotted as symbols: Northern Spur (cross), And NE (circle), eastern
Merrett PNe (triangle), the Eastern Shelf (square), M32 (star), And
VIII (diamond), and the G1 clump (star). }
\end{figure}
\begin{figure}
\includegraphics[width=84mm,bb=0 8 288 485]{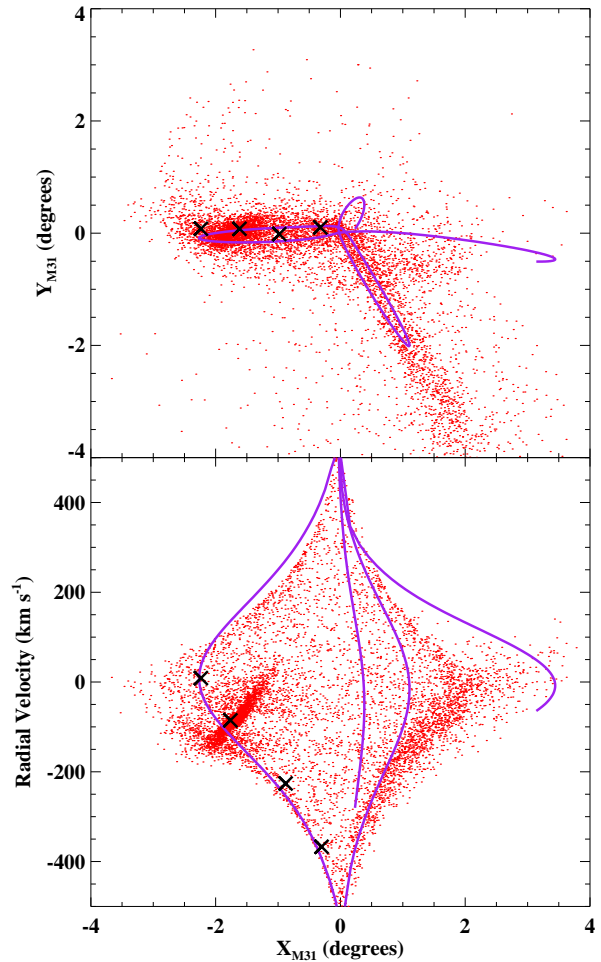}
\caption{
\label{fig.merrettfit}
Best-fitting orbit passing through the Merrett et al.\ (2003, 2004)
PNe.  In this plot we have changed to their M31-aligned
coordinate system to allow easier comparison with their figures.
$X_{M31}$ increases along M31's major axis to the SW, and $Y_{M31}$
along its minor axis to the NE.  The four crosses in each panel show
approximate points along the narrow-velocity-dispersion
group of PNe identified in \protect\citet{merrett04}.}
\end{figure}
{\it Linear continuation}: \citet{mcconnachie03} observed two fields
along the linear projection of the southern stream on the other side
of the disk.  Even though there is no obvious projection seen in
surface brightness maps, they obtained red giant color-magnitude
diagrams for these fields consistent with a stream component. We
assign a position to this feature using the center of Field 13 and
errors of $0.2^\circ$.  Since the projected orbit tends to bend
substantially at pericenter, we find it is very difficult to fit this
feature as the progenitor itself, in agreement with the conclusions of
\citet{ibata04} and \citet{font04}.  However, it is possible that the
stream scatters enough debris to the side when it passes through pericenter 
to produce the stars seen by McConnachie et al.

{\it NGC 205 and nearby arc}: NGC 205 is close to the projection of
the stream to the other side of M31.  However, its velocity has the
opposite sign from that of the stream. If it is the progenitor, it
must lie nearly at the {\em next} pericenter, i.e. at phase $F_p \sim
2.0$.  Because of its location north of M31, it is difficult to fit an
orbit with this constraint.  The stellar arc through the center of NGC
205 \citep{mcconnachie04} produces the same difficulties with the
orbit as NGC 205 itself.  In addition, the bluer color of the red
giants in NGC 205 and the arc indicate a lower metallicity compared to
the stream, again arguing against a connection with the stream.

{\it M32}: M31's other prominent nearby satellite galaxy is an obvious
candidate for the origin of the stream, and was proposed as such as
soon as the stream was discovered \citep{ibata01}.  M32 lies almost
directly on the path of the stream. However, as with NGC 205, its
radial velocity has the wrong sign for it to lie within the southern
stream.  Hence it too would have to be at the next pericenter ($F_p
\sim 2.0$).  Interestingly, on orbital grounds alone, it is not
possible to rule out this possibility.  We find an acceptable fit,
even though the observational constraints on this fit are more
numerous and more precise than for any other candidate (since the
position, distance, and radial velocity are all fairly well known).
The angle of the second loop in this fit is not very robust, and could
accomodate a wide range of position angles ranging from 0 to
$60^\circ$, which encompasses a range of observed structures, although
the best-fitting orbit, shown in Fig.~\ref{fig.m32fit}, does not result
in an obvious match to any in particular.  The progenitor mass before
last pericenter required to produce the length of the stream is about
$M_s = 2.5 \times 10^9 \msun$.  However, it then must lose much of its
mass to the forward and backward streams in order to produce the
observed luminosity of the stream.  \citet{mateo98} gives the {\it
  current} mass of M32 as $2.1 \times 10^9 \msun$.  Within the
uncertainties, the masses are probably consistent.  From studying the
isophote ellipticity of M32, \citet{choi02} inferred that it is on a
highly eccentric orbit with $R_{peri} < 1.7 \kpc$, which is consistent
with our best-fitting orbit.  However, these authors also inferred that
M32 is most likely moving eastward; this is not consistent with our
orbit, though it is difficult to evaluate the reliability of their
argument.  There are other difficulties with our specific orbit: it
creates a plume in the northward direction that has not been reported
observationally, and it is not clear that the stream forward of M32
can match the observed structure.  The metallicity distribution of M32
is complex, and it is not clear whether or not it would be consistent
with that of the stream.  Nevertheless, on balance of the evidence,
M32 continues to be a possibility for the progenitor of the stream.

\begin{figure}
\includegraphics[width=84mm,bb=0 8 504 485]{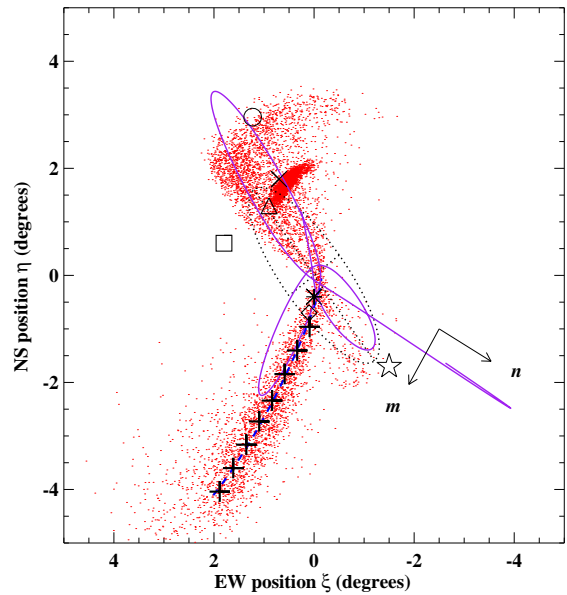}
\caption{
\label{fig.spurfit}
Best-fitting orbit passing through the Northern Spur, and the particle
distribution in the corresponding N-body simulation.  Crosses, orbits,
and dots have the same meaning as in Fig.~\ref{fig.tiltorbit}.
Object symbols are the same as in Fig.~\ref{fig.m32fit}.  The
progenitor is visible as the dense elongated clump near the NE end of
M31.}
\end{figure}
{\it Merrett \ea PNe}: \citet{merrett03} surveyed the
disk of M31 for PNe and found a number that had
velocities inconsistent with the disk kinematics.  Almost by
definition, most of them move in a direction opposite to the local
rotation of the disk, though a few move in the same direction but with
larger speeds.  \citet{merrett04} pointed out a subset of these on the
northeast side that occupy a narrow region in position-velocity space.
We can easily fit this group as the progenitor of the stream.
(Merrett et al.\ also fit an orbit connecting this group to the stream
using their simpler potential, but without demonstrating agreement
with the detailed properties of the stream.)  Choosing the position
and velocity to lie within the northeast group, we find an excellent
fit with the initial conditions given in Table~\ref{table.streamfits}.
Fig.~\ref{fig.merrettfit} shows the positions and velocities
obtained from an N-body simulation using this orbit, using a
coordinate system aligned along the major axis of M31.  These can be
compared to fig.~3 of \citet{merrett04}.  Interestingly, there are a
significant number of N-body particles occupying larger velocities at
a given position than the main concentration, but none with smaller
velocities; a similar pattern is seen in \citet{merrett04}.  This
feature is thus a prime candidate for the progenitor of the stream.

{\it Northern Spur}: The star-count maps of M31 in \citet{ferguson02}
show a faint clump of stars at the NE end of M31, called the Northern
Spur.  We estimate the position of the Spur from the image in
\citet{zucker04}, which also is suggestive of a loop of emission
extending from the Spur.  The red giant branch color is similar to
that of that of the stream \citep{ferguson02}.  This feature is a
prominent candidate for the continuation of the stream
\citep{merrett03}.

Some radial velocity observations exist in the Spur region.  From
spectroscopy of red giants, \citet{ferguson04} claimed a velocity of
$\tsim 150 \kms$, significantly different from that of the disk in
this region, suggesting it is a distinct object.  \citet{merrett04}
found an overdensity of PNe in this region.  The
velocities of some of these are consistent with the disk velocity in
this region, some have larger velocities, and some have the wrong sign
altogether.  These authors inferred that the Spur represents a warp in
the disk of M31.  The possible contamination from the disk and the
presence of velocities of both signs suggest that if the progenitor
lies in the Spur region, its velocity cannot be reliably constrained
at present.

If we impose no velocity constraints, we find an excellent fit to the
position of this feature, as seen in Fig.~\ref{fig.spurfit}.  The
predicted location is on the near side of M31, and the progenitor is
outward bound from pericenter, giving a negative radial velocity.
Thus slight differences in its color-magnitude diagram in
\citet{ferguson02}, as compared to that of the stream, could possibly
be explained by its smaller distance.  One possible problem is the
faint luminosity of the Spur; from the area and surface brightess
discussed in \citet{ferguson02}, we estimate its $V$-band luminosity
is only $8 \times 10^6 \lsun$, less than that of the stream.  The
simulation shown in the figure produces a fairly dense group of
particles at the position of the Spur, and would probably need to be
adapted to produce a fainter, lower-surface-brightness core.

If we instead assume that the velocity is positive, as suggested by
the majority of the velocity data, the progenitor must be moving
inwards once again.  In this case it is more difficult to fit a model
to the Spur.  Our best fit is formally ruled out, though this conclusion
is dependent on the assumed errors for the Spur's properties
(we assume a velocity $200 \pm 30 \kms$ and a positional error
of $0.2^\circ$).  

{\it And NE}: This low surface brightness enhancement far out along
the major axis of M31 was discussed by \citet{zucker04}.  We find it
is quite easy to make the orbit pass through this object.  However,
this solution is objectionable on other grounds: there is no visible
connection to the stream, despite the low surface brightness of M31 at
these projected radii, and the luminosity is probably too low for it
to be an intact progenitor.  

{\it Eastern shelf}: \citet{ferguson02} discussed a diffuse surface brightness
enhancement on the eastern side of M31, which contained
high-metallicity red giants judging by their color.  This feature
is now known as the ``Eastern Shelf''.  This color may be
consistent with that of the Northern Spur and of the stream itself.
With the position we have assigned to this feature, it is not easy to
find an orbit where this feature is the progenitor itself, despite
the weakness of the observational constraints.  This
conclusion differs from that in \citet{font04} because the greater
strength of their potential, and their use of an orbit that
follows the stream, affect the direction of the second loop.  The
low surface brightness of this feature also argues against it being
the progenitor.  It might represent debris from the
continuation of the stream, though for none of the orbits we tried
did the debris reach the position angle of this feature.  

{\it And VIII}: This is another group of objects with abnormal radial
velocities in the M31 disk \citep{morrison03}, in this case a
combination of PNe, globular clusters, and H~I detections.  This
feature lies nearly across the stream and has a velocity consistent
with it, but it is extended for $\tsim 1^\circ$ transverse to the
stream, which is not expected for a progenitor moving along the
stream.  Hence if it is the progenitor, it has probably passed through
pericenter already.  Incorporating this condition as a constraint, we
find that it is very difficult to fit an orbit to this feature.

{\it G1 Clump}: \citet{ferguson02} identified a diffuse surface
brightness enhancement in the vicinity of the globular cluster G1,
which they called the ``G1 clump''.  We did not attempt to fit this as
the progenitor of the southern stream.  By virtue of its position,
this would have to lie at $F_p \approx 2.5$.  This would make for a
very massive progenitor. The G1 clump, however, is less luminous and
thus less massive than the stream, and it is not clear that the
progenitor mass in excess of these two structures can be hidden.  In
addition, the color-magnitude diagram of the clump differs from that
of the southern stream.  However, it is at least suggestive that in a 
number of our N-body simulations of other possible progenitors, there is
a clump or shell-like feature in roughly the vicinity of this feature,
where the forward section of the stream turns around (see
Figs.~\ref{fig.m32fit} and \ref{fig.spurfit}).

\vskip+0.75cm 
In summary, the preferred mass and orbital phase of the progenitor
indicate there is probably a significant amount of structure near M31
that is physically connected to the stream but ahead of it in its
orbit, amounting to $\gtrsim 10^9 \msun$ in total.  We can probably
rule out most of the objects on the southwestern half of M31 as
progenitors, with the possible exception of M32.  We can fit
reasonable orbits through several features in the northeastern half:
the Northern Spur, And NE, and the Merrett \ea PNe.
N-body simulations following these orbits produce a spray of debris as
well as a number of discrete-looking features; there is some
resemblance between this debris and the structure seen in the
observations, even though we do not have a detailed match.  We have
made significant approximations in the fitting and simulation process,
among them the neglect of dynamical friction, the assumption of a
static fixed potential, and the neglect of the satellite's
self-gravity in deriving the best-fitting orbit.  Thus we have not tried
to refine the match to the observed features any further.  However, by
predicting the distance and velocity of each feature, we hope to speed
the process of confirming or rejecting a connection between
the various faint features and the southern stream.
\section{Summary}
In this paper and Paper I, we have attempted to construct a model for
the interaction of M31 and the southern stream.  We constructed a
simple analytic model for the potential, and found its best fit
parameters by fitting to a number of observational constraints.  We
then combined this galaxy fit with a fit to the observed path of the
southern stream.  We found the best fit parameters as a function of
the orbital phase of the progenitor.  By considering the effect of the
progenitor's mass on the observed properties of the stream, we derived
an N-body model of the interaction that fits current observations
reasonably well.  We tested the orbital fits against various objects
and surface brightness features near M31, ruling out some as the progenitor
and making predictions for the kinematic properties of the others.

In comparison to the stream from the Sagittarius dwarf galaxy that
surrounds the Milky Way, there is currently much more leeway in
fitting the Andromeda stream.  The progenitor of the former stream is
known, and its proper motion has been measured.  Distance errors for
this stream are a factor of 10 smaller than for M31.  In addition,
this stream has been followed for nearly a full circle around the
Galaxy, as opposed to the Andromeda stream which spans less than half
a radial period.  Thus, this stream is currently a much better probe
of its parent halo's properties than in the case of M31 and its
southern stream.  In this paper, we have found that the orbital phase
of the progenitor is a major source of uncertainty in the orbital
parameters.  If we can identify the progenitor location, we can
immediately put much stronger constraints on the orbital properties of
the stream.  We can then get a better understanding of the mass
distribution in M31, calculate the properties of the progenitor
satellite, untangle the possible relationships of the various surface
brightness features, and perhaps find signatures of the passing
satellite in the M31 disk.

Our specific conclusions are as follows:
\vskip-0.75cm
\begin{itemize}
\item The stream follows a path in phase space significantly different
from the orbit of the progenitor satellite.  Allowing for this effect
decreases the estimated apocenter and 
increases the estimated gravitational force in the M31 halo.
\item The narrowness and surface brightness are consistent with a
young stream, resulting from the last pericentric passage of its
progenitor.  Larger ages are ruled out by the narrowness and high
luminosity of the stream.
\item With the current observational uncertainties, the stream itself
does not add much constraining power to the galaxy model.  Our best
combined fit for the mass of M31 within 125 kpc is $(7.4 \pm 1.2) \times
10^{11} \msun$, only slightly larger than in our best-fitting model from
Paper I.
\item 
If the progenitor satellite has orbital phase $F_p = 1.0$ 
(at the next pericenter beyond the stream location),
our best fit to the orbit of the progenitor has radial period
0.8 Gyr, apocenter 63 kpc, and pericenter 2.8 kpc.
However, $F_p$, which determines the size and time scales of
the orbit, is poorly determined.  
We find it likely that the satellite or its disrupted remnants 
lie beyond next pericenter.
\item The dynamical mass of the progenitor at the previous pericentric
passage was $M_s \approx 10^9 \msun$, within a factor 3.
Its mass was not heavily dominated by dark matter; otherwise, the
surface brightness of the stream would be less than is observed.
\item Depending on the exact properties of the progenitor, significant
amounts of debris from the progenitor may lie outside the currently
observed width of the stream.  These stars could contaminate the
metallicity distribution measured in off-stream fields, but
can be detected since they share the velocity 
and distance signatures of the stream.
\item Most of our orbital fits would place the continuation 
of the stream past next pericenter and to the NE of M31's center.  
We can fit reasonable orbits through several observed features, 
among them the Northern Spur, And NE, M31, and a feature in
the planetary nebula distribution found by Merrett et al.
The latter feature seems to be the most plausible candidate 
for the progenitor of the stream.
\end{itemize}
\section*{ACKNOWLEDGMENTS}
We thank Tom Quinn, Joachim Stadel, James Wadsley, and Josh Barnes for
the use of their simulation and analysis codes.  Andreea Font, Kathryn
Johnston, Steve Majewski, and Mike Rich graciously conveyed their
revised stream model prior to publication.  We thank Dave Balam, Chris
Pritchet, and Andreea Font for helpful conversations.  Research
support for JJG, MF and AB comes from the Natural Sciences and
Engineering Research Council (Canada) through the Discovery and the
Collaborative Research Opportunities grants.  AB would also like to
acknowledge support from the Leverhulme Trust (UK) in the form of the
Leverhulme Visiting Professorship.  PG is supported by NSF grant
AST-0307966.  He is grateful to the HIA/DAO/NRC staff for graciously
hosting his 2002-03 Herzberg fellowship during which time this
collaborative project was conceived.

\label{lastpage}


\begin{thebibliography}{}
\bibitem[Binney \& Tremaine(1987)]{binney87} Binney, J.~\&
  Tremaine, S.\ 1987, ``Galactic Dynamics'',
  Princeton University Press, Princeton, NJ
\bibitem[Choi, Guhathakurta, \& Johnston(2002)]{choi02} Choi, 
P.~I., Guhathakurta, P., \& Johnston, K.~V.\ 2002, AJ, 124, 310 
\bibitem[Durrell, Harris, \& Pritchet(2001)]{durrell01} Durrell, 
P.~R., Harris, W.~E., \& Pritchet, C.~J.\ 2001, AJ, 121, 2557 
\bibitem[Durrell, Harris, \& Pritchet(2004)]{durrell04} Durrell, 
P.~R., Harris, W.~E., \& Pritchet, C.~J.\ 2004, AJ, 128, 260 
\bibitem[Ferguson et~al.(2002)]{ferguson02} Ferguson, A.~M.~N., Irwin, M.~J.,
    Ibata, R.~A., Lewis, G.~F., \& Tanvir, N.~R.\ 2002, AJ, 124, 1452
\bibitem[Ferguson et~al.(2004)]{ferguson04} Ferguson, A.,
Chapman, S., Ibata, R., Irwin, M., Lewis, G., McConnachie, A. 2004,
in Proc.\ of ESO-Workshop ``Planetary Nebulae beyond the Milky Way'',
eds.\ J. R. Walsh, L. Stanghellini, Springer-Verlag (astro-ph/0408058)
\bibitem[Font et~al.(2005)]{font04} Font, A., Johnston, K.~V., Guhathakurta,
    P., Majewski, S.~R., \& Rich, R.~M.\ 2005, AJ, in press (astro-ph/0406146)
\bibitem[Geehan et al.(2005)]{geehan05}
  Geehan, J. J., Fardal, M. A., Babul, A., Guhathakurta, P.
  2005, MNRAS, in press (Paper I)
\bibitem[Guhathakurta et~al.(2005)]{raja05} Guhathakurta, P., Rich, R. M.,
    Reitzel, D.~B., Cooper, M.~C., Gilbert, K., Majewski, S.~R., Ostheimer,
    J.~C., Geha, M. C., Johnston, K. V., \& Patterson, R. J. 2005, AJ,
    in press (astro-ph/0406145)
\bibitem[Ibata et~al.(2001a)]{ibata01} Ibata, R., Irwin, M.~J., Ferguson,
    A.~M.~N., Lewis, G., \& Tanvir, N.\ 2001a, Nature, 412, 49
\bibitem[Ibata et~al.(2001b)]{ibata01sag} Ibata, R., Lewis, G. F., 
    Irwin, M. J., Totten, E., \& Quinn, T.  2001b, ApJ, 551, 294
\bibitem[Ibata et~al.(2004)]{ibata04} Ibata, R., Chapman, S., Ferguson,
    A.~M.~N., Irwin, M., Lewis, G., \& McConnachie, A.\ 
    2004, MNRAS, 351, 117
\bibitem[Johnston(1998)]{johnston98} 
Johnston, K.~V.\ 1998, ApJ, 495, 297
\bibitem[Johnston, Sackett, \& Bullock(2001)]{johnston01}
Johnston, K.~V., Sackett, P.~D., \& Bullock, J.~S. 2001, ApJ, 557, 137
\bibitem[Johnston, Law, \& Majewski(2005)]{johnston05}
Johnston, K.~V., Law, D. R., \& Majewski, S. R. 
2005, ApJ, 619, 800
\bibitem[Landau \& Lifschitz(1976)]{landau} Landau, L. D., \&  Lifschitz, E. M.
1976, ``Mechanics'', Pergamon, Oxford
\bibitem[Law, Johnston, \& Majewski(2005)]{law05} 
Law, D. R., Johnston, K.~V., \& Majewski, S. R. 
2005, ApJ, 619, 807  
\bibitem[Mateo(1998)]{mateo98} Mateo, M.~L.\ 
   1998, ARAA, 36, 435 
\bibitem[Mart{\'{\i}}nez-Delgado et al.(2004)]{martinez04} Mart{\'{\i}}nez-Delgado, D., G{\' 
o}mez-Flechoso, M.~{\' A}., Aparicio, A., \& Carrera, R.\ 2004, ApJ, 601, 
242 
\bibitem[McConnachie et~al.(2003)]{mcconnachie03} McConnachie, A.~W., 
 Irwin, M.~J., Ibata, R.~A., Ferguson, A.~M.~N., Lewis, G.~F., \& Tanvir, N.\ 
 2003, MNRAS, 343, 1335
\bibitem[McConnachie et al.(2004)]{mcconnachie04} McConnachie, A.~W., 
Irwin, M.~J., Lewis, G.~F., Ibata, R.~A., Chapman, S.~C., Ferguson, 
A.~M.~N., \& Tanvir, N.~R.\ 2004, MNRAS, 351, L94 
\bibitem[Merrett et al.(2003)]{merrett03} Merrett, H.~R., et al.\ 
2003, MNRAS, 346, L62 
\bibitem[Merrett et al.(2004)]{merrett04} Merrett, H., et~al.\
2004, in Proc.\ of ESO-Workshop ``Planetary Nebulae beyond the
Milky Way'', eds.\ J. R. Walsh and L. Stanghellini, Springer-Verlag
(astro-ph/0407331)
\bibitem[Morrison et~al.(2003)]{morrison03} Morrison, H.~L., Harding, P.,
Hurley-Keller, D., \& Jacoby, G. 2003, ApJ, 596, L183
\bibitem[Peebles et al.(2001)]{peebles01} 
Peebles, P.~J.~E., Phelps, S.~D., Shaya, E.~J., \& Tully, R.~B.\ 2001, 
ApJ, 554, 104 
\bibitem[De Rijcke et~al.(2005)]{derijcke05}
De Rijcke, S., Michielsen, D., Dejonghe, H., Zeilinger, W. W., \& Hau, G. K. T.
  2005, A\&A, 438, 491
\bibitem[Stadel(2001)]{stadel01}
Stadel, J. 2001, Ph.D.\ Thesis, University of Washington, Seattle, WA, USA
\bibitem[Stanek \& Garnavich(1998)]{stanek98} Stanek, K. Z., \& 
Garnavich, P. M. 1998, ApJ, 503, 131L
\bibitem[Taylor \& Babul(2001)]{taylor01} Taylor, J.~E., \& 
Babul, A.\ 2001, ApJ, 559, 716 
\bibitem[Totten \& Irwin(1998)]{totten98} Totten, E.~J.~\& 
Irwin, M.~J.\ 1998, MNRAS, 294, 1 
\bibitem[de Vaucouleurs et~al.(1991)]{devauc91}
de Vaucouleurs, G., de Vaucouleurs, A., Corwin, H., Buta, R., 
  Paturel, G., \& Fouque, P., 1991, ``Third Reference Catalogue 
  of Bright Galaxies'', Springer-Verlag, Berlin/Heidelberg/New York
\bibitem[Wadsley, Stadel, \& Quinn(2004)]{wadsley04} Wadsley,
J.~W., Stadel, J., \& Quinn, T.\ 2004, New Astronomy, 9, 137
\bibitem[Zucker et al.(2004)]{zucker04} Zucker, D.~B., et al.\ 
2004, ApJ, 612, L117
\end{thebibliography}
\end{document}